\documentclass[useAMS, usenatbib, usegraphicx]{mn2e}
\usepackage{aasmacros}
\usepackage{ifpdf}

\title[Linking Subhalo Accretion and Satellite Orbits]{The link between galactic satellite orbits and subhalo
accretion} 
\author[M. R. Lovell, V. R. Eke, C. S. Frenk, and A. Jenkins]{Mark
R. Lovell$^{1}$, Vincent R. Eke$^{1}$, Carlos S. Frenk$^{1}$, and
Adrian Jenkins$^1$\\ 
$^{1}$Institute for Computational Cosmology, Durham University, South
Road, Durham, DH1 3LE} 
\begin{document}

\date{Accepted 2011 January 17. Received 2010 December 9; in original form 2010 July 28} 

\pagerange{\pageref{firstpage}--\pageref{lastpage}} \pubyear{2010}

\maketitle

\label{first page}

\begin{abstract}
We calculate the orbital angular momentum of dark matter subhaloes in
the Aquarius simulations of cold dark matter (CDM) galactic haloes. We
calculate the orientation of their angular momentum relative to that
of the spin vector of their host halo and find a variety of different
configurations. All six Aquarius haloes contain statistically
significant populations of subhalo orbits that are aligned with the
main halo spin. All haloes possess a population of subhaloes that rotates
in the same direction as the main halo and three of them possess, in
addition, a population that rotates in the opposite direction. These
configurations arise from the filamentary accretion of subhaloes. 
Quasi-planar distributions of coherently
rotating satellites, such as those inferred in the Milky Way and other
galaxies, arise naturally in simulations of a $\Lambda$CDM universe.
\end{abstract}

\begin{keywords}
galaxies: formation - dark matter 
\end{keywords}

\section{Introduction}
\label{In}

It has been known for several decades that the 11 `classical'
satellites of the Milky Way define a thin plane around the Galaxy
\citep{ly76}. Some of the faint satellites recently 
discovered in the Sloan Digital Sky Survey
\citep[SDSS;][]{yo00,wi05a,wi05b,sa06,zu06a,zu06b,ir07, wa07,be08} also appear to have an anisotropic
distribution reminiscent of that of the classical 11 \citep{me09}.
The presence of such a `disc-of-satellites' suggests a common plane
of rotation in the Milky Way. Measurements of proper motions, which
are now possible for some of the satellites, can be used to constrain
the nature of any systemic rotation \citep{me08, lu10}. 

In tandem with these observational developments, advances in
computational cosmology now make it possible to simulate galactic
haloes with sufficient resolution to probe the properties of satellites
and investigate the origin of their flattened configuration.
N-body simulations from cold dark matter (CDM) initial conditions show
that a large number of accreted haloes survive to the present, making
up a population of `subhaloes' of the `main halo,' some of which could
host the satellites.

The observations, however, suggest a complex formation
history. First, the number of satellites identified so far is much
smaller than the number of dark subhaloes in the simulations, giving
rise to the so-called `missing satellite problem'
\citep{mo99a,kl99}. Secondly, the thin ring around the sky
delineated by the classical satellites contrasts with the distribution
of subhaloes in the simulations which is triaxial
\citep{li05,ze05}. Thirdly, the inferred angular momentum vectors of
the majority of the classical satellites in the Milky Way point
towards a patch on the sky of diameter no greater than $30^{\circ}$,
which has led \cite{me08,me09} to argue that the observed satellites
cannot have formed in cold dark matter subhaloes. In contrast to the
Milky Way, NGC~5084 appears to have a population of satellites
orbiting in the opposite sense to the galaxy \citep{ca97}.

The combination of `missing satellites', an anisotropic distribution
and coherent orbits is sometimes viewed as a challenge to the CDM
model \citep[e.g.][]{mo99a,me08}. However, a number of studies using
semi-analytic modelling and hydrodynamic simulations have shown that a
relatively small satellite population is a natural outcome of galaxy
formation in the CDM cosmology
\citep[e.g][]{Kauffmann93,Bullock_00,Benson02_sats,Somerville02, ko09, mu09, bu10, co10, li10, ma10, wad10}.
The simulations show that satellites form only in a small fraction of
subhaloes which turn out to be those that had the most massive
progenitors at the time of accretion \citep{li05}.  Furthermore,
disc-like subhalo configurations are seen to form in $\Lambda$CDM
cosmological simulations
\citep{ka05,ze05,li07,li08,li09}. Such systems appear to be related 
to the preferential accretion of haloes along the filaments of the cosmic
web. Haloes tend to fall along the central spines of filaments, so
that the range of trajectories, and thus orbits, that they acquire
when they enter a halo is restricted \citep{li09}. 

\cite{sh06}, \cite{wa06} and \cite{li09} confirmed the conclusion of \cite{li05} 
that satellite accretion is a highly anisotropic process and found in their simulations a
significant population of subhaloes that
co-rotated with the spin of their hosts. However, \cite{sh06}
simulated galaxy cluster haloes, not galactic haloes;
\cite{wa06} also focused on cluster haloes except for one example of a galaxy halo which, however,
had only moderate resolution (a minimum subhalo mass of
$m_{\mathrm{min}}=5.7\times10^{7}\mathrm{M}_{\odot}$).  The largest sample of galaxy
halo simulations so far is that of \cite{li09}. They analysed 436 haloes
but were only able to resolve subhaloes of mass $m_{\mathrm{min}}\geq 2.76\times10^{9}\mathrm{M}_{\odot}$.

In this study, we analyse the state-of-the-art, high resolution
simulations of six galactic haloes of mass
$\sim1\times10^{12}\mathrm{M}_{\odot}$ of the Aquarius project
\citep{sp08b}. These simulations resolve subhaloes of mass 
exceeding $\sim10^{5}\mathrm{M}_{\odot}$.  We calculate the angular momentum of
subhaloes, and use the results to interpret the Milky Way data. The
paper is organised as follows. In Section \ref{SC} we briefly describe
the Aquarius project and the analysis performed for this paper. The
results follow in Section~\ref{Re} and our conclusions in
Section~\ref{Co}.

\section{ß The simulations}
\label{SC}

\begin{table*}
  \begin{tabular}{ccccrccccc}
    \hline
    Name & $m_{\mathrm{p}}$ [$\mathrm{M}_{\odot}$] & $r_{200}$ [kpc] & $M_{200}$ [$\mathrm{M}_{\odot}$] & $n_{\mathrm{s}}$ & $\lambda$ & $c^{*}_{\mathrm{NFW}}$ & $z_{\mathrm{form}}$ & $q$ & $p$\\
    \hline
    Aq-A1 & $1.712\times10^{3}$ & 245.76 & $2.523\times10^{12}$ & 197484 & - & 16.11 & 1.93 & -  & -  \\
    Aq-A2 & $1.370\times10^{4}$ & 245.88 & $2.524\times10^{12}$ & 30177 & $0.027$ & 16.19 & 1.93 & 0.866  & 0.687  \\
    Aq-A3 & $4.911\times10^{4}$ & 245.64 & $2.524\times10^{12}$ & 9489 & - & 16.35 & 1.93 & 0.862  & 0.688  \\
    Aq-A4 & $3.929\times10^{5}$ & 245.70 & $2.524\times10^{12}$ & 1411 & - & 16.21 & 1.93 & 0.844  & 0.700  \\
    Aq-A5 & $3.143\times10^{6}$ & 246.37 & $2.541\times10^{12}$ & 246  & - & 16.04 & 1.93 & 0.830  & 0.685  \\
    \hline
    Aq-B2 & $6.447\times10^{3}$ & 187.70 & $1.045\times10^{12}$ & 31050 & $0.022$ & 9.72 & 1.39 & 0.820* & 0.839* \\
    \hline
    Aq-C2 & $1.399\times10^{4}$ & 242.82 & $2.248\times10^{12}$ & 24628 & $0.020$ & 15.21 & 2.23 & 0.711* & 0.770* \\
    \hline
    Aq-D2 & $1.397\times10^{4}$ & 242.85 & $2.519\times10^{12}$ & 36006 & $0.012$ & 9.37 & 1.51 & 0.846* & 0.901* \\
    \hline
    Aq-E2 & $9.593\times10^{3}$ & 212.28 & $1.548\times10^{12}$ & 30372 & $0.017$ & 8.26 & 2.26 & 0.898* & 0.674* \\
    \hline
    Aq-F2 & $6.776\times10^{3}$ & 209.21 & $1.517\times10^{12}$ & 35041 & $0.050$ & 9.82 & 0.55 & 0.700\dag & 0.866\dag \\
    \hline
  \end{tabular}
  \caption{Selected parameters of the Aquarius simulations used in
this paper. The simulation name encodes the halo label (Aq-A, B, and
so on) and the numerical resolution level (1 to 5, hereafter L1, L2, L3, L4, L5). $m_{\mathrm{p}}$ is the particle mass,
$r_{200}$ the radius of the sphere of density 200 times the critical
density, $M_{200}$ the halo mass within $r_{200}$, $n_{\mathrm{s}}$
the number of subhaloes within the main halo, $\lambda$ the
spin parameter as determined by \citet{bo10}, and $q$, $p$ the halo
shape axis ratios $b/a$ and $c/b$ respectively (Vera-Ciro et al., in
preparation). The axes are defined as $a \geq b \geq c$ for ellipsoids
determined using the method of \citet{al06}. Values with * or \dag~~
superscripts were calculated for haloes at resolution levels L4 or L3
respectively. As the smallest subhaloes determined by
\textsc{subfind} contain 20 particles, the minimum subhalo mass in
each simulation is $20m_{\mathrm{p}}$.} 

  \label{table1}
\end{table*}

\begin{figure}
  \includegraphics[scale=0.44]{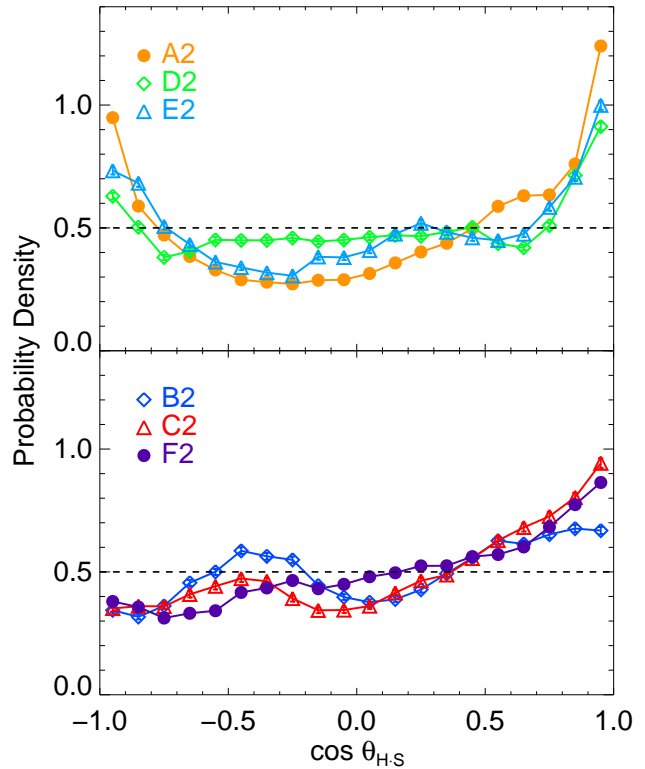}
  \caption{The
  distribution of cos~$\theta_{\mathrm{H\cdot S}}$ for the six Aquarius haloes
  at resolution level L2, where $\theta_{\mathrm{H\cdot S}}$ is the angle between the main halo spin and subhalo orbit vectors. The six are separated into two panels
  according to whether or not they exhibit an antiparallel tail
  greater than 0.5. \emph{Top panel:} results for Aq-A2
  (orange), Aq-D2 (green), and Aq-E2 (light blue). \emph{Bottom
  panel:} as above, but for Aq-B2 (blue), Aq-C2 (red), and
  Aq-F2 (purple). In both cases, the dashed line corresponds to an
  isotropic distribution.} \label{MOT1}
\end{figure}

The Aquarius project is a set of dark matter simulations containing
haloes similar in size and environment to those of the Milky Way; each
one has been run from $z=127$ to $z=0$. There are six different haloes
(Aq-A - Aq-F), each of which has been resimulated at at least two levels
of resolution (L1, the highest, down to L5, the
lowest). They were performed using the \textsc{gadget-3} code
\citep{sp08b}. In all cases, the resimulations at different resolutions show
remarkable convergence in the positions and internal properties of
subhaloes. This project has already yielded several interesting
results, including a study of the near-universality of halo density
profiles \citep{sp08b,na10}, predictions for the $\gamma$-ray signal
from annihilating dark matter in the galactic halo~\citep{sp08a} and
for direct dark matter detection experiments~\citep{Vogelsberger09}.
 A summary of key parameter values for each simulation is given in
Table~\ref{table1}. 

It is important for this study to establish that the sample
of six Aquarius haloes can be considered at least approximately
representative of the population of Milky-Way mass haloes as a whole.
The Aquarius haloes are all drawn from the same parent cosmological
simulation, and it is possible to address this issue directly for
several properties. The spins, concentrations and formation histories
of the Aquarius haloes are compared to the parent population in
\citet{bo10}.  Broadly speaking, the properties of the Aquarius haloes
span the expected range for the population as a whole. We give the
values of the halo spin, concentration and formation redshift, defined
as the redshift when half the halo mass is assembled, in
Table~\ref{table1}. Also in the table we list the shape axis ratios
for the haloes, approximating them as ellipsoids.  The axis ratios are
taken from Vera-Ciro et al. (in preparation) and calculated for ellipsoids
which are determined by applying the iterative method of \cite{al06}
to the haloes with the substructure removed (actually to the `main
halo', defined below).  The six haloes show a range of shapes and are
typical for $\Lambda$CDM haloes \citep{al06,be07}.

The halo membership of each particle is determined using the
friends-of-friends (FOF) algorithm \citep{da85}. The particles in each
FOF group are, in turn, assigned to self-bound structures using the
\textsc{subfind} code \citep{sp01a}. We call the largest of
these self-bound substructures the main halo, and the remainder we
call subhaloes.  A small proportion ($<1$ per cent) of the
particles within the FOF group are found to form a `fuzz' that is not
gravitationally bound to any other object; they are not considered any
further.

Our primary aim is to determine the orientations of dark matter
subhalo orbits in the Aquarius simulations and compare the results
with data for galactic satellites. We calculate the `main halo spin',
defined as the sum of the angular momenta of all main halo particles
about their centre-of-mass. For each subhalo, we calculate the
`subhalo orbital spin', defined as the vector associated with the
angular momentum of each subhalo about the centre of the main halo.  We then
calculate the cosine of the angle, $\theta_{\mathrm{H\cdot S}}$, between the
main halo spin vector and the subhalo orbit vector for every subhalo
associated with that main halo. These subhaloes are tracked back to
the initial conditions in order to investigate the origin of the
patterns that we find.

\begin{figure}
  \includegraphics[scale=0.44]{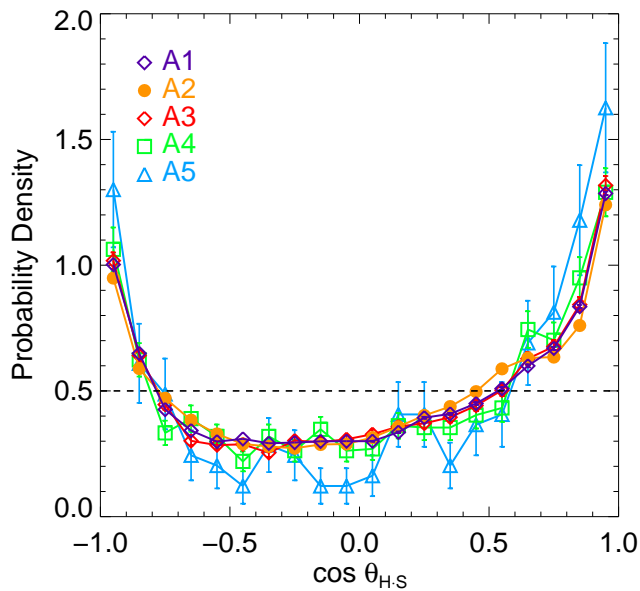}
  \caption{The distribution of cos~$\theta_{\mathrm{H\cdot S}}$ for all the
resolution levels of Aq-A. Aq-A1 contains 197484 subhaloes (purple), Aq-A2 30177 (orange),
Aq-A3 9489 (red), Aq-A4 1411 (green), and Aq-A5 246 (light blue). The error bars denote
Poisson uncertainties.} 
  \label{MOA532} 
\end{figure}

\section{Results}
\label{Re}

\begin{figure}
  \includegraphics[scale=0.44]{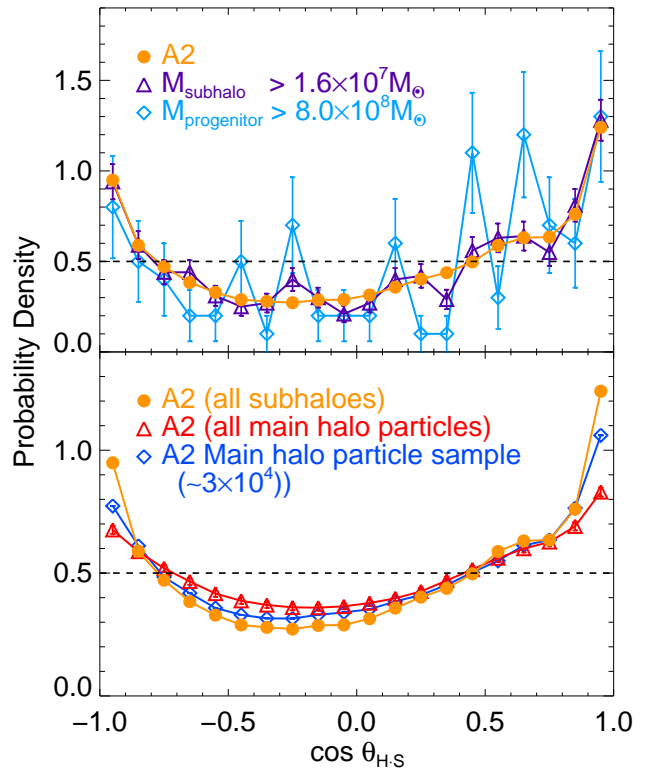} \caption{The
  distribution of cos~$\theta_{\mathrm{H\cdot S}}$ for different
  populations. \emph{Top panel:} comparison of the 1000 most massive
  subhaloes at $z=0$ (purple) with the 100 subhaloes that have the
  most massive progenitors (light blue), and the entire population of
  Fig.~\ref{MOA532} (orange). \emph{Bottom panel:} comparison
  of the cos~$\theta_{\mathrm{H\cdot S}}$ distribution for subhaloes with two
  populations of main halo particles: a sample of $3\times10^{4}$
  selected to have the same radial distribution as the main halo 
  (blue) and the full population (red).} 
  \label{MassBins}
\end{figure}

We first describe our calculation of the angular momentum
distributions of various populations and then investigate their
origin.

\subsection{Angular momentum distributions of subhaloes}

We compute cos~$\theta_{\mathrm{H\cdot S}}$ for each of the six L2 haloes at
$z=0$ as described above, and plot the results in Fig.~\ref{MOT1} as
a probability density; an isotropic distribution of angular momenta in
this plot corresponds to a horizontal line at $0.5$.

All the haloes show a statistically significant bias for
subhalo orbits to be aligned (parallel) to the rotation of the parent
main halo, as found by \citet{sh06} and \citet{wa06}. The average
fraction of corotating subhaloes in the Aquarius haloes is 57 per cent, with
a narrow range between 54 per cent and 61 per cent. This is consistent with the
average of 59 per cent quoted by  \citet{wa06}. This result
is a natural outcome of tidal torque theory \citep{ho51,wh84} when the
primordial dark matter protohaloes exert torques on one another,
inducing net spins as they condense.

We also find significant numbers of nearly
\emph{anti}-parallel orbits in three of our haloes. Specifically,
haloes Aq-A2, Aq-D2, and Aq-E2 show a significant proportion
of subhalo orbits in the $-1.0<$cos~$\theta_{\mathrm{H\cdot
S}}<-0.9$ bin (9.5 per cent, 6.3 per cent, and 7.3 per cent
respectively where 5 per cent would be expected for a random
distribution - the Poisson errors on our L2 measurements are
negligible), whilst Aq-B2, Aq-C2, and Aq-F2 do not. We have
separated the haloes into two panels according to this property.
We find an antiparallel excess in three out of six of our
haloes, whereas \citet{wa06} only have one such halo out of their
sample of nine. Adopting the same binning as \cite{wa06} does not
change our result.  With such small halo samples it is unclear whether
this particular result is consistent or inconsistent between the two
studies.

To test if our results are robust to changes in resolution, we repeat
this calculation for the five different resolution levels of
the Aq-A halo (Fig.~\ref{MOA532}). We see that Aq-A1 together with Aq-A3, Aq-A4, and Aq-A5 has an angular momentum
distribution broadly of the same form as Aq-A2, with increasing
noise as the resolution decreases because of the smaller number of
subhaloes. Each resolution level is dominated by a different subhalo
mass; the minimum subhalo mass in Aq-A5 is
$\sim10^{7}\mathrm{M}_{\odot}$, while in Aq-A1 it is three
orders of magnitude smaller. We find a similar degree of
convergence with numerical resolution for haloes Aq-B through to Aq-F.

In Fig.~\ref{MassBins} we probe the orientation of the angular
momentum vector of different populations. In the top panel, we compare
the distribution for the 1000 largest subhaloes at the final redshift
(particle number $>1222$, equivalent to subhalo mass of
$1.7\times10^{7}\mathrm{M}_{\odot}$) with that the 100 subhaloes present at
$z=0$ that had the most massive progenitors and that of the entire
halo population. The most massive progenitor is defined as the \textsc{subfind}
halo in the merger tree that contained the largest number of particles
over the entire history of the simulation. This mass is very close to
the mass that the subhalo had at the time it fell into the main
halo. It is these subhaloes that are most likely to host satellite
galaxies, according to \citet{li09}. Of the subhaloes that had the 100 largest progenitors, all bar 6 are among the top 1000 most massive subhaloes at redshift zero. 
The distributions of cos~$\theta_{\mathrm{H\cdot S}}$ for all three populations of subhaloes are consistent within the errors. 

To establish whether the angular momentum orientation of the
subhalo population is special, in the lower panel of
Fig.~\ref{MassBins} we compare subhaloes in Aq-A2 with
particles from the main halo.  We create a special sample of
halo particles with the same radial distribution as the
subhaloes. This is made by first defining a set of about 30 radial
bins between the halo centre and the virial radius. The halo subsample
is produced by first noting how many subhaloes lie in a particular
bin, and then randomly selecting the same number of halo particles
from the that same bin. This is always possible as the number of halo
particles in any bin exceeds the corresponding number of
subhaloes.
We compare this particle sample's distribution
of cos~$\theta_{\mathrm{H\cdot S}}$ with that for the Aq-A2
subhaloes and for the entire set of main halo particles. The three
distributions are statistically inconsistent with each other. The
subhalo population has a larger fraction of aligned and antialigned
members, with the radially selected subsample being
intermediate between the subhaloes and the halo particles as a
whole. Although even the latter has a non-uniform distribution of
angular momenta cosines, it is significantly flatter than that of
other two populations.  This suggests that the accretion mechanism
that supplies subhaloes (of all masses) is somewhat different from the
mechanism by which halo particles are accreted, or that the evolution
of subhaloes differs from that of halo particles.

\begin{figure*}
$\begin{array}{c@{\hspace{-0cm}}c@{\hspace{-0cm}}}
        \includegraphics[angle=-90,width=0.50\textwidth]{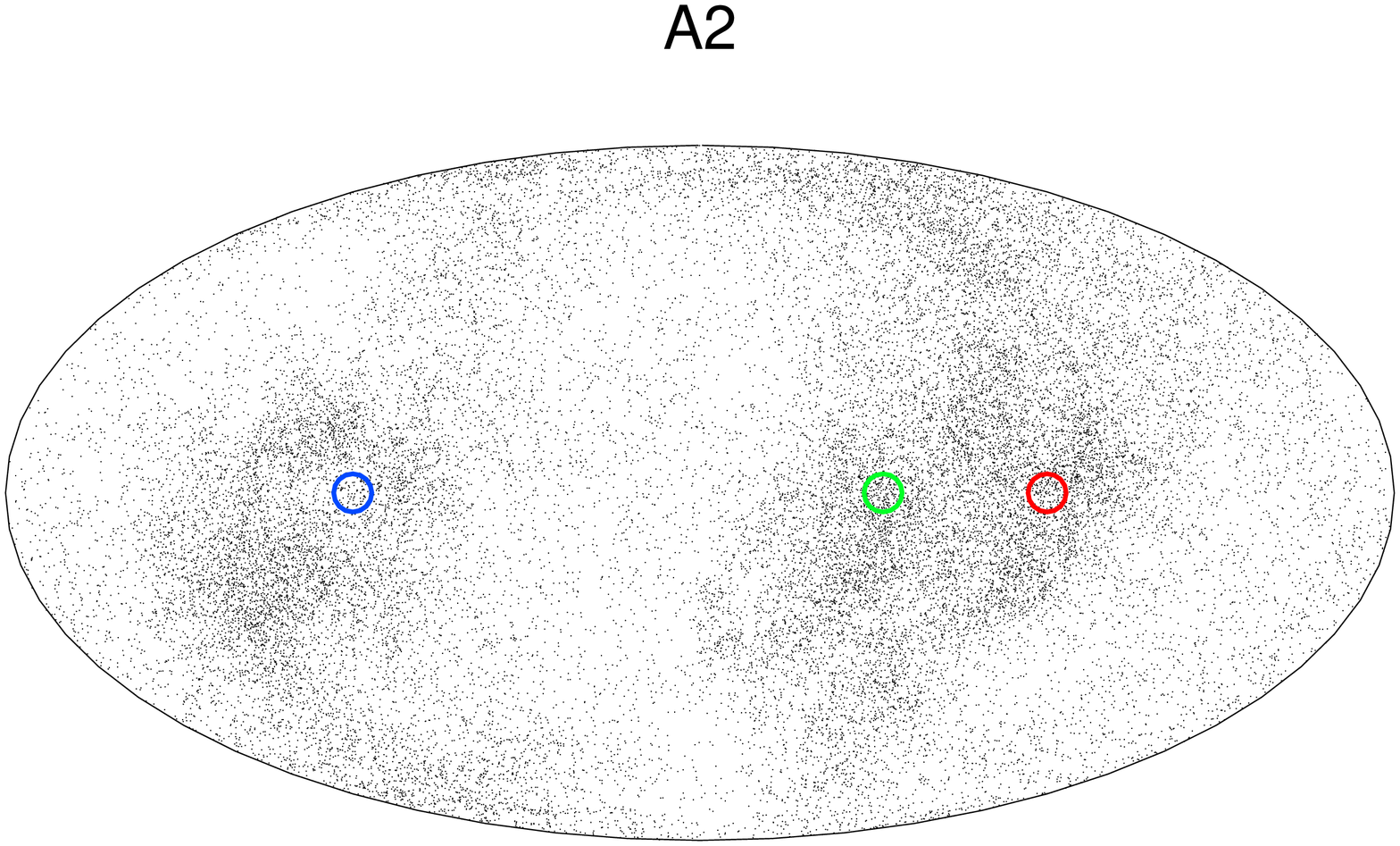} &
        \includegraphics[angle=-90,width=0.50\textwidth]{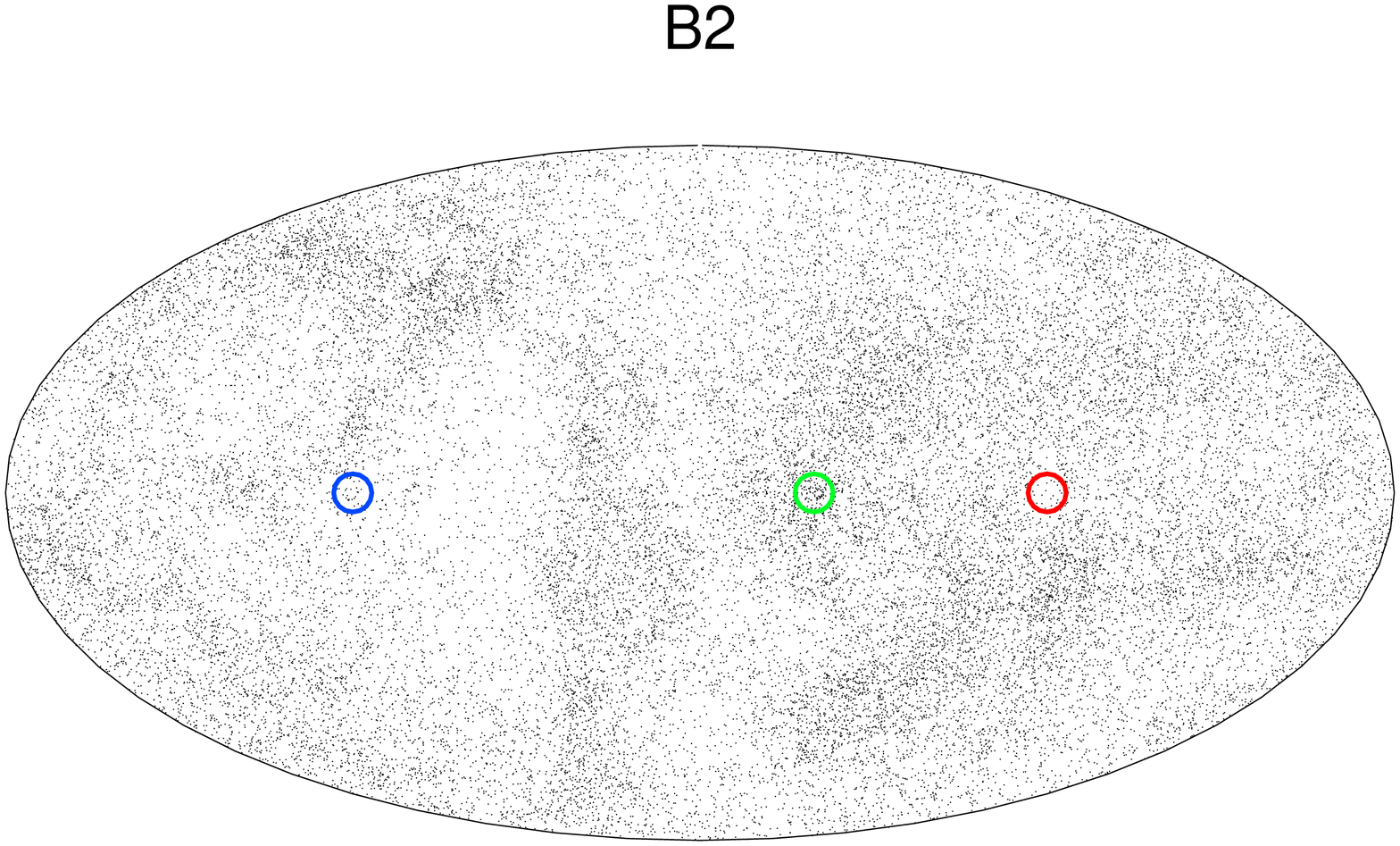}\\
	\includegraphics[angle=-90,width=0.50\textwidth]{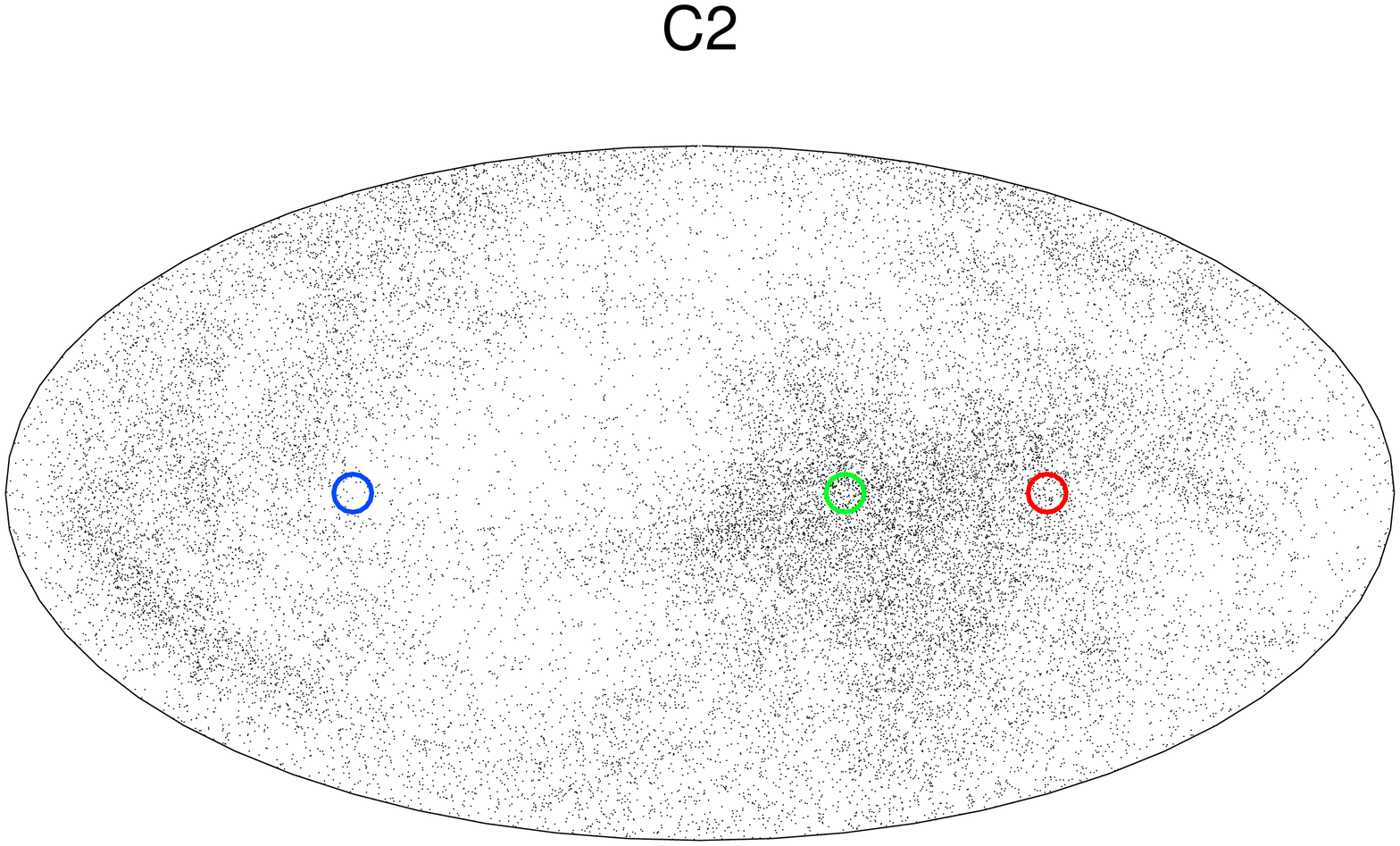} &
        \includegraphics[angle=-90,width=0.50\textwidth]{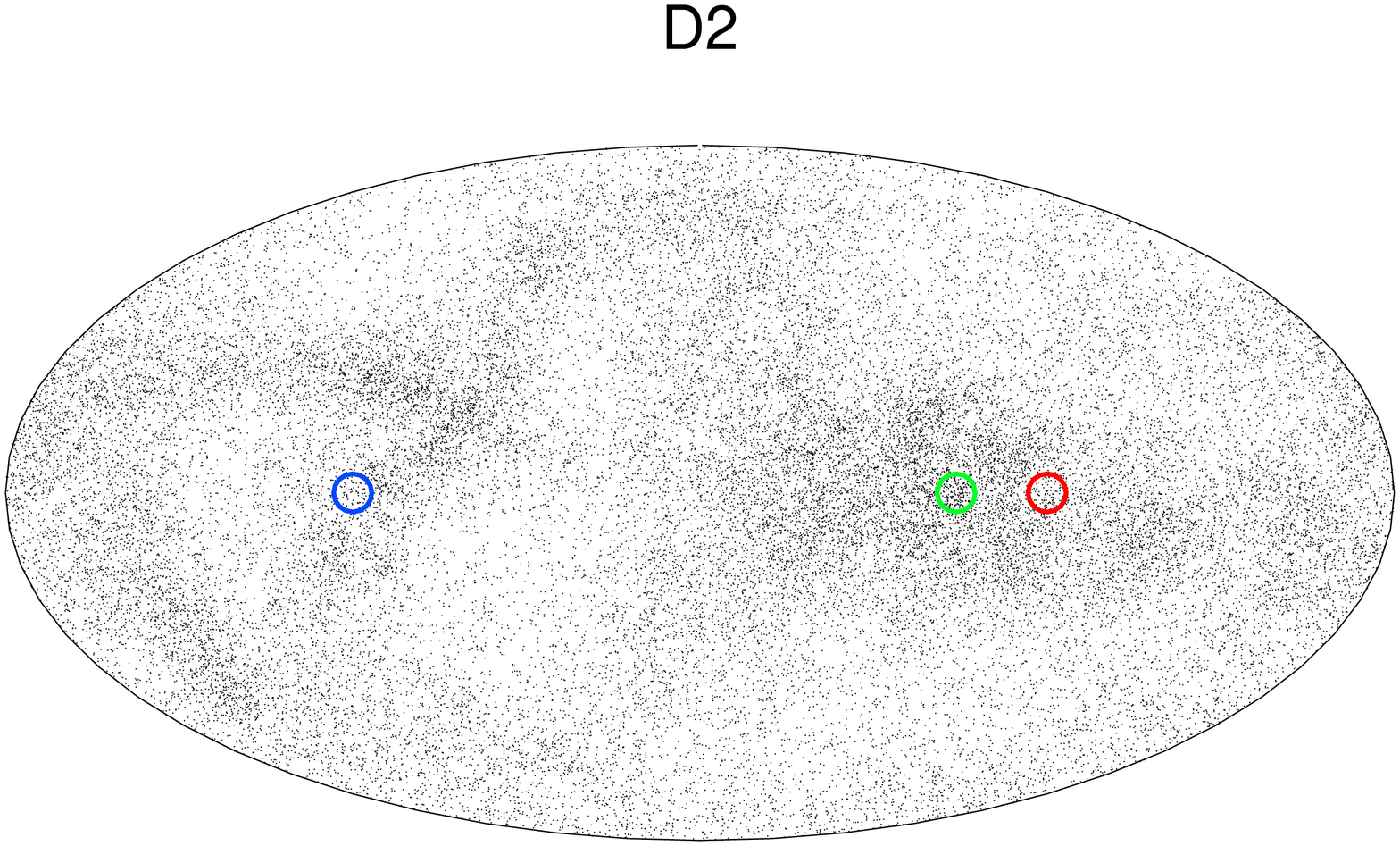}\\
	\includegraphics[angle=-90,width=0.50\textwidth]{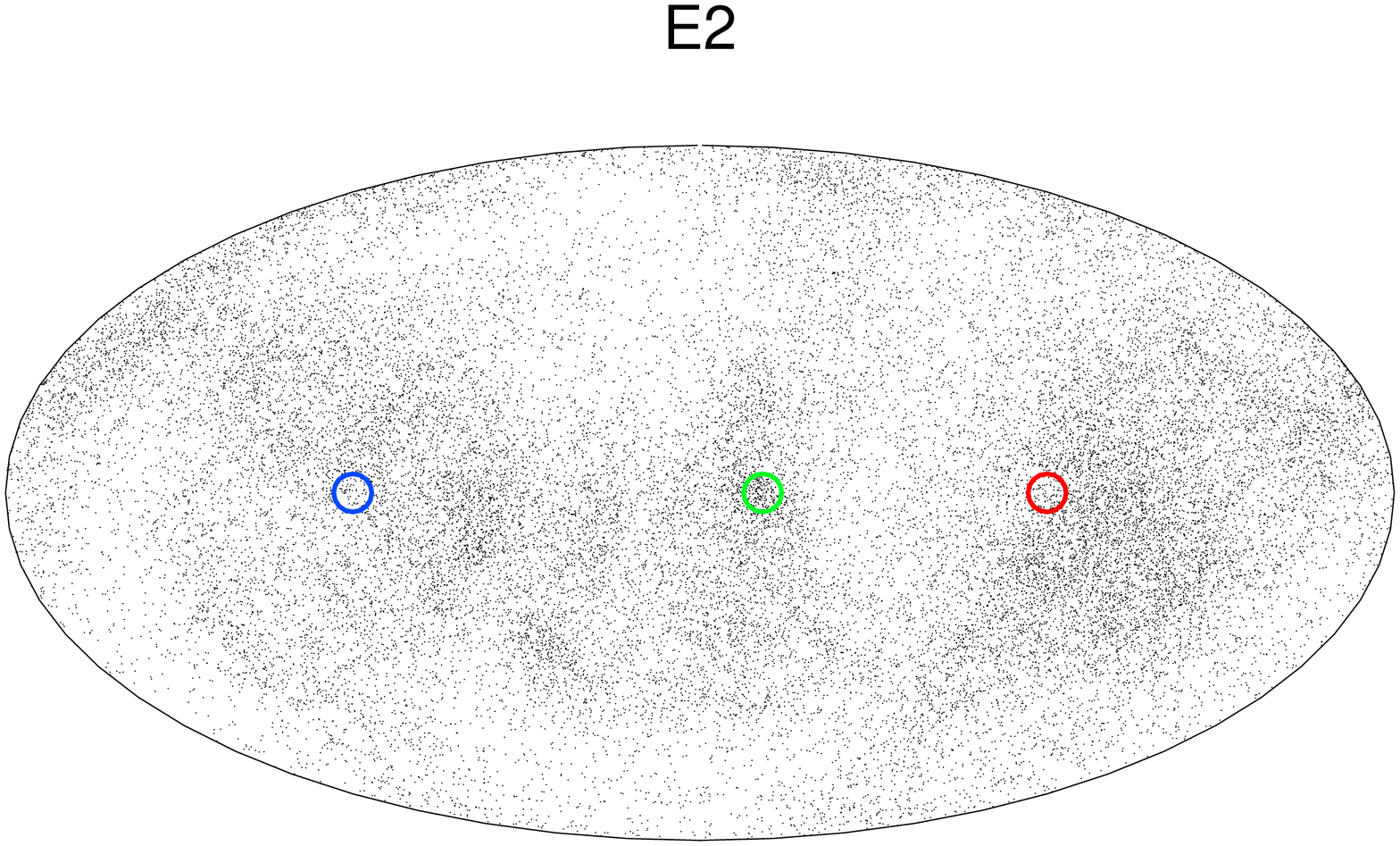} &
        \includegraphics[angle=-90,width=0.50\textwidth]{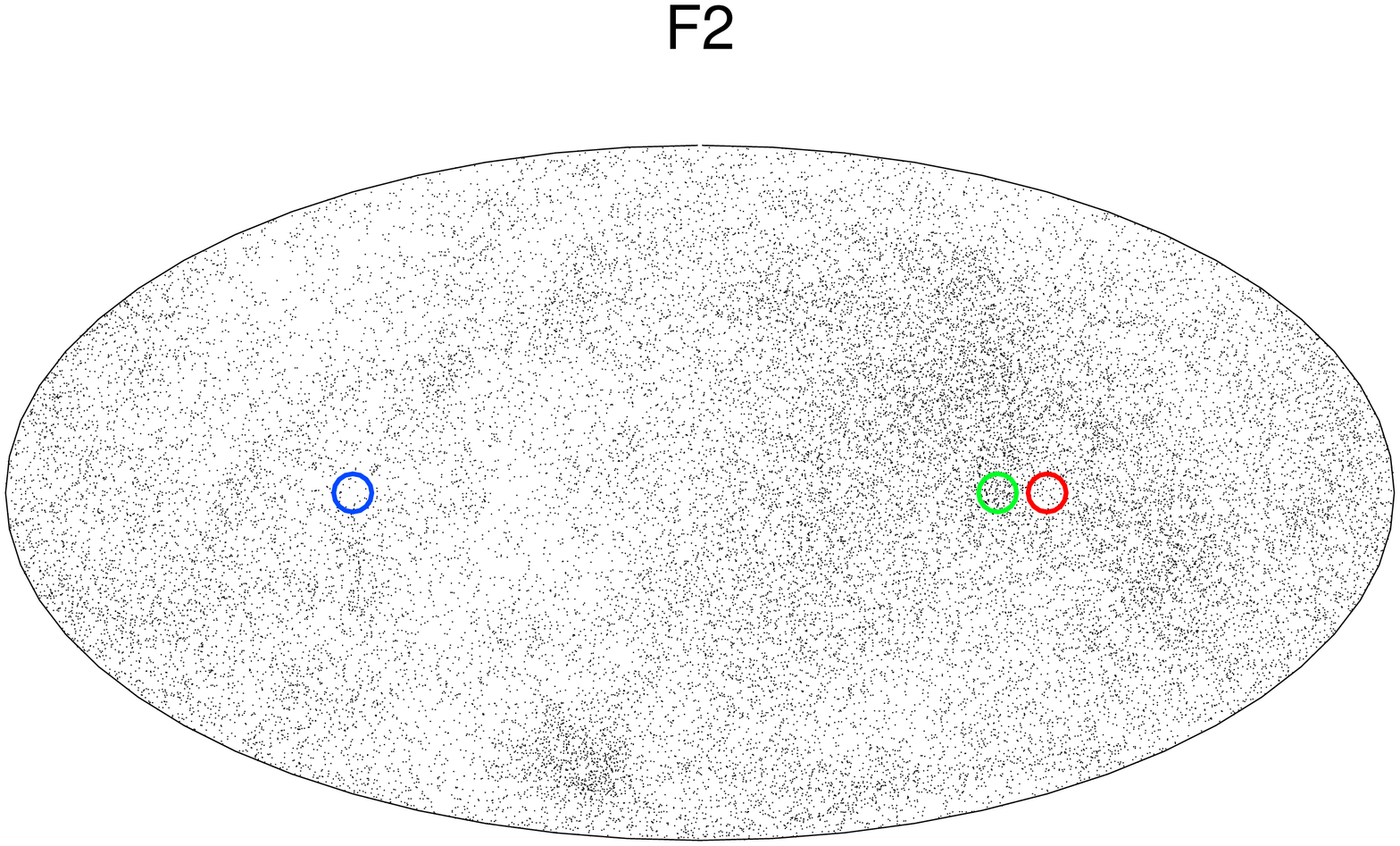}\\
\end{array}$
 \caption{Mollweide projections of the directions of the angular
 momentum vectors of all subhaloes in the L2 simulations. The red circle
 shows the direction of the main halo spin, blue the main halo spin
 antipole, and green the densest collection of vectors after 
 smoothing. The maps have been rotated such that all three circles lie on
 the equator, with the main halo spin and its antipole lying 
 90$^{\circ}$ either side of the centre and the green circle in
 between. Thus, a subhalo of $\theta_{\mathrm{H\cdot S}}$$=0^{\circ}$ will map to
 the red circle, and one of $\theta_{\mathrm{H\cdot S}}$$=90^{\circ}$ to either the plot boundary or a point on the north-south bisector.}
\label{MapOrbits}
\end{figure*}

To investigate the orientation of the orbital spins in more detail, we
plot the angular momentum vectors of each
subhalo on an all-sky Mollweide projection, one for each halo at
resolution L2.  Each map displayed here was divided into  $\sim45000$ pixels, with angular width
$\sim1^{\circ}$, and smoothed with a Gaussian beam of FWHM
$10^{\circ}$ using Healpix routines
\citep{Go04}. We identify the pixel with the highest density
after smoothing, and call this the `densest point vector'. The pre-smoothing maps for all six L2 haloes are displayed in Fig~\ref{MapOrbits}. The main
halo spin vector is marked in red, its antipole in blue, and the
densest point vector in green.

\begin{figure*}
$\begin{array}{c@{\hspace{-0cm}}c@{\hspace{-0cm}}}
        \includegraphics[angle=-90,width=0.50\textwidth]{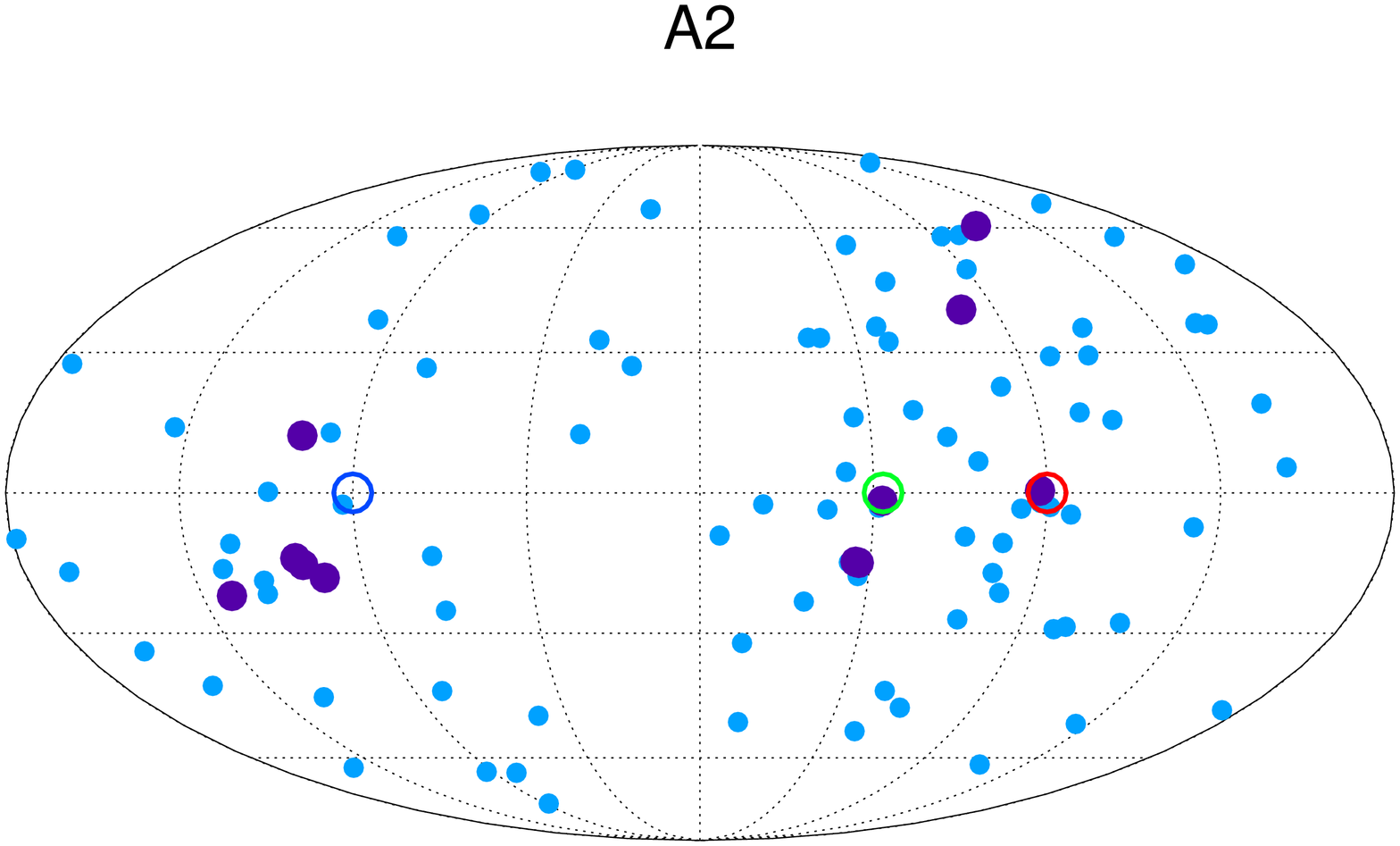} &
        \includegraphics[angle=-90,width=0.50\textwidth]{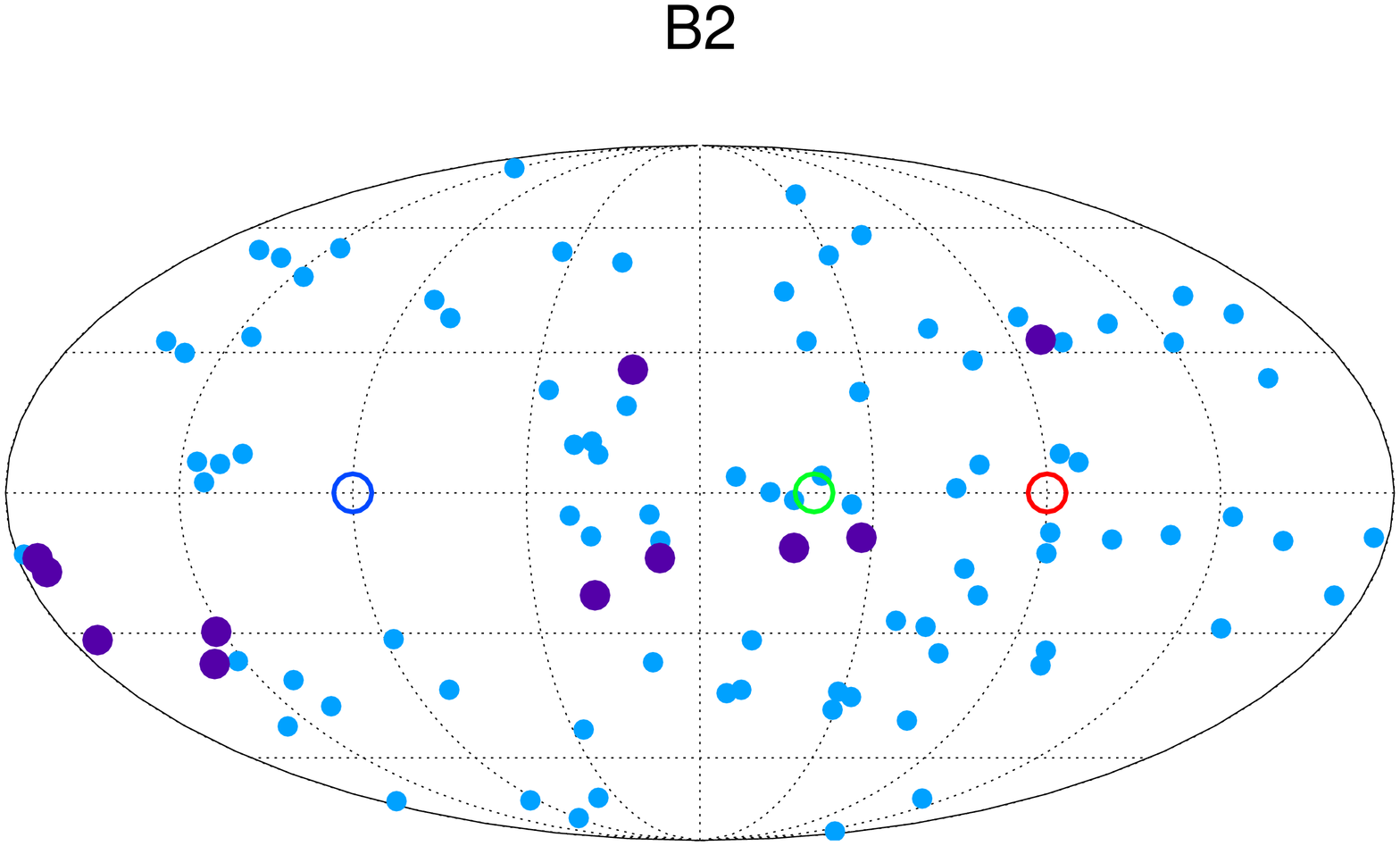}\\
	\includegraphics[angle=-90,width=0.50\textwidth]{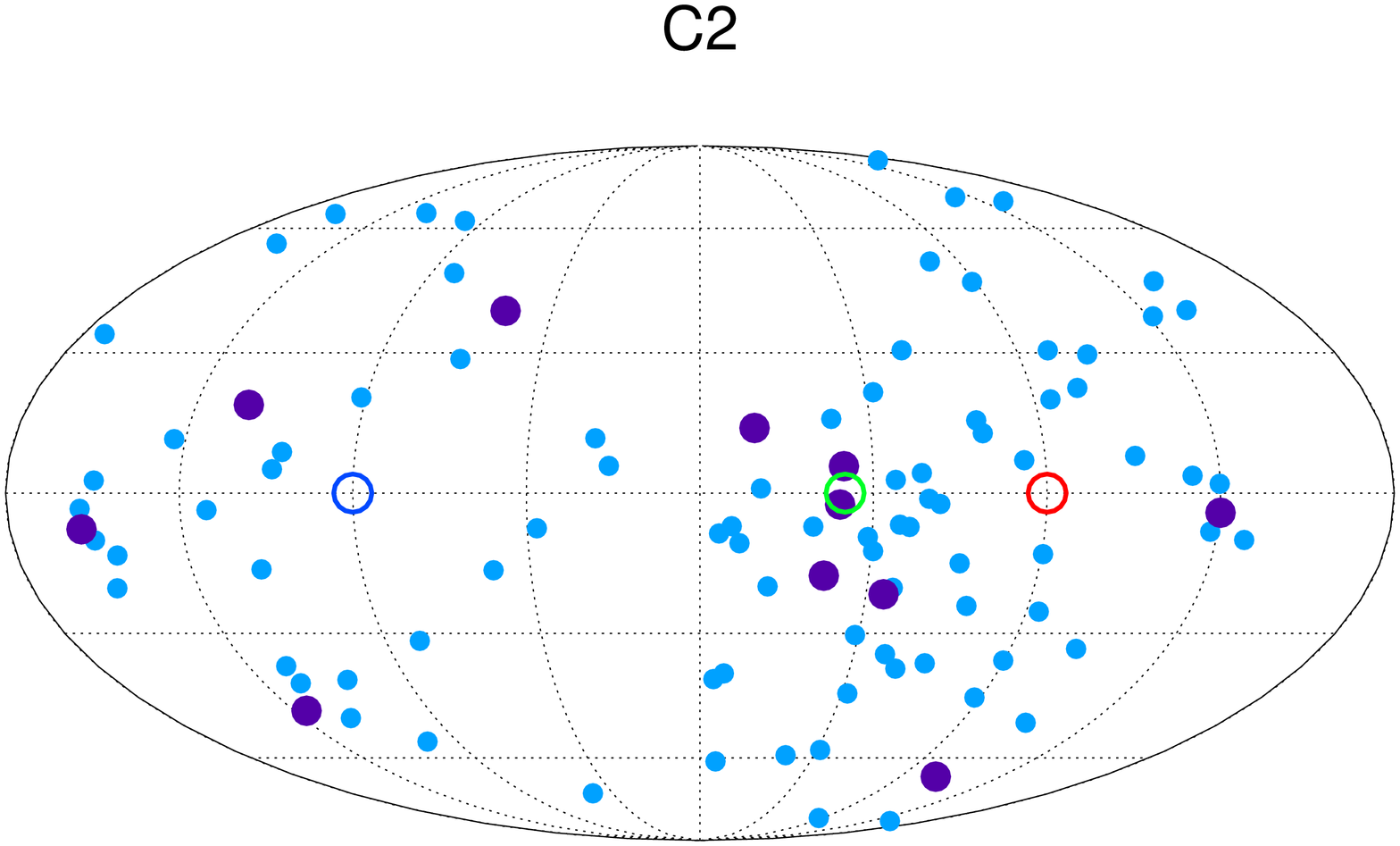} &
        \includegraphics[angle=-90,width=0.50\textwidth]{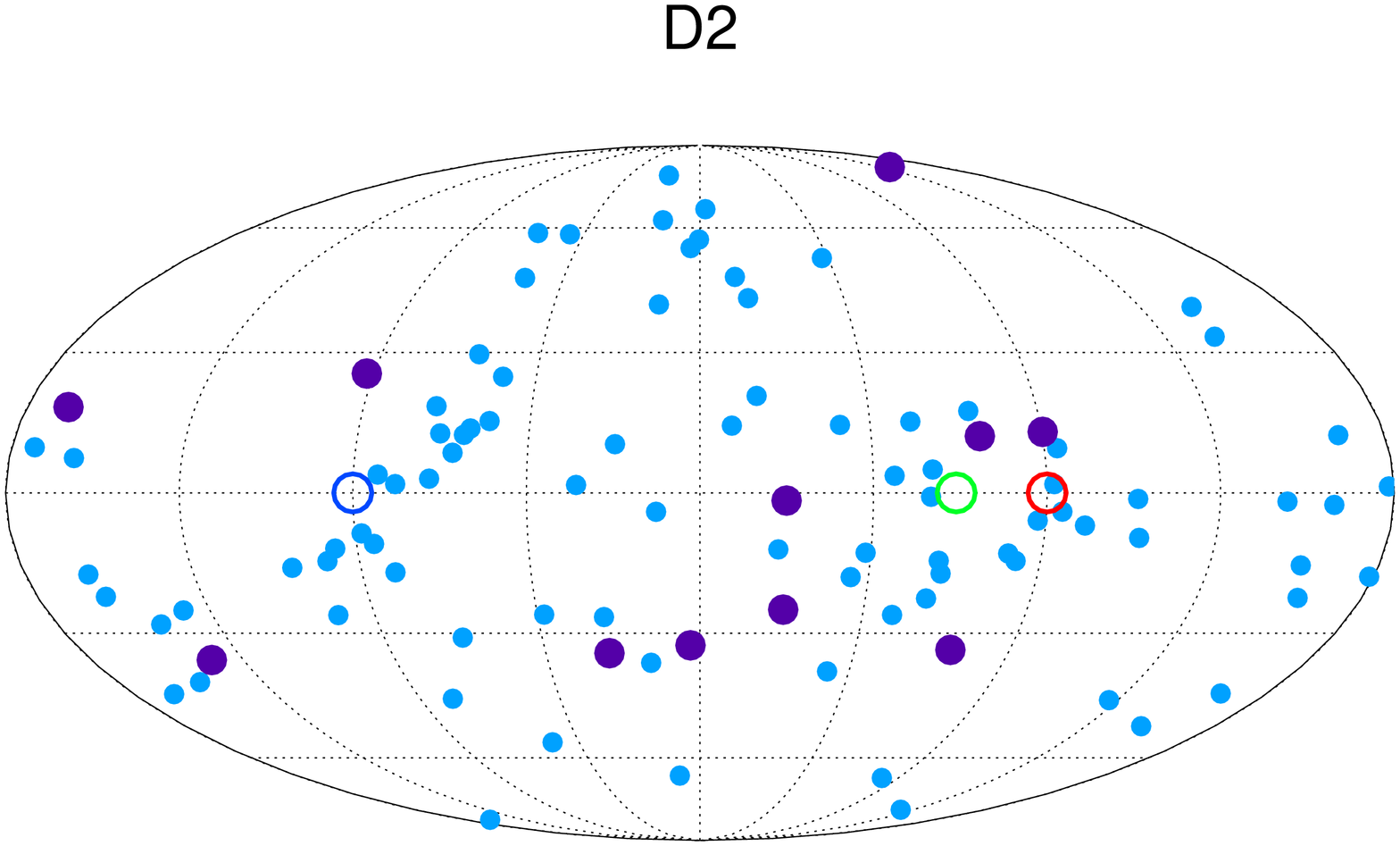}\\
	\includegraphics[angle=-90,width=0.50\textwidth]{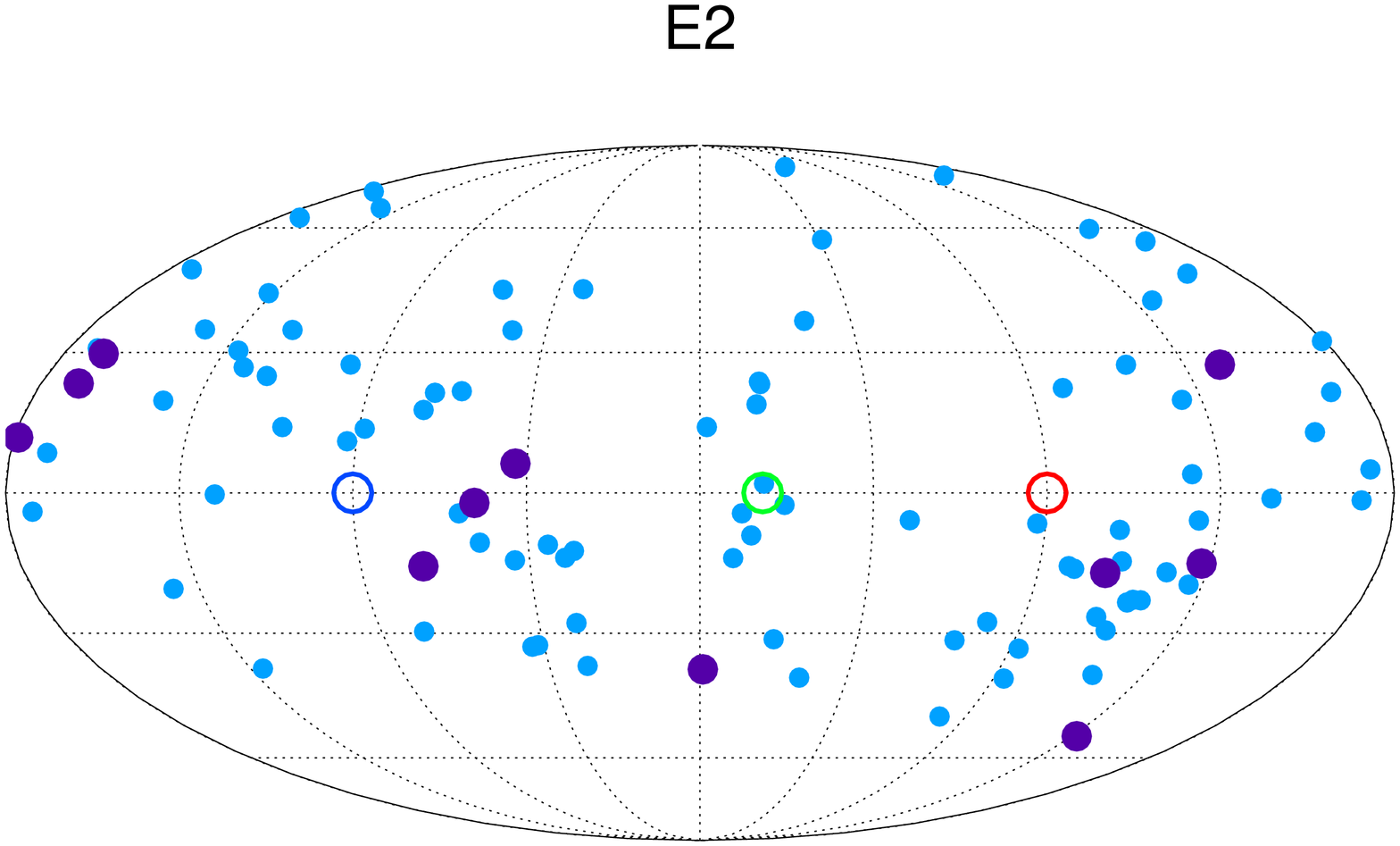} &
        \includegraphics[angle=-90,width=0.50\textwidth]{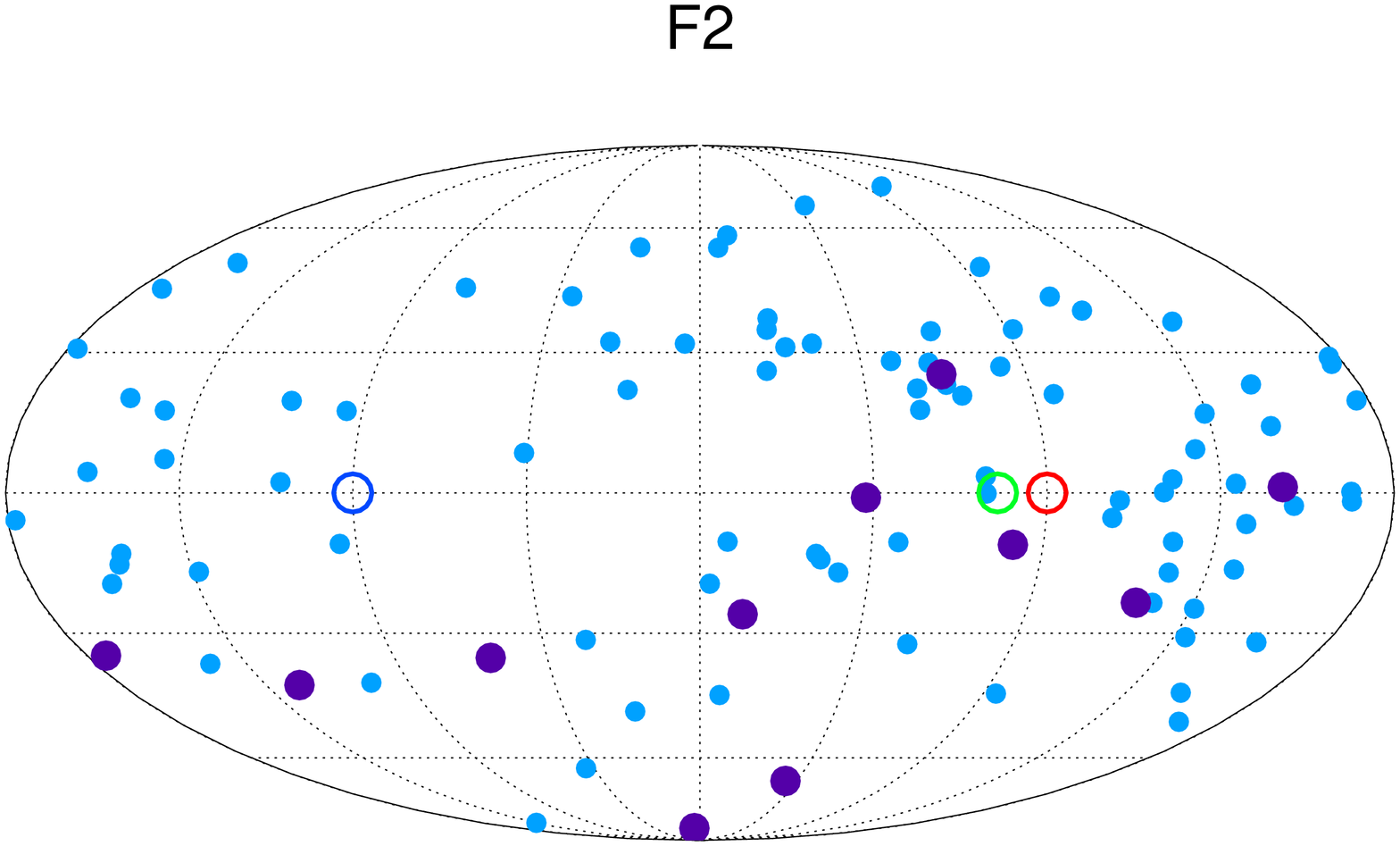}\\
\end{array}$
\caption{Mollweide projections of the directions of the angular
 momentum vectors of subhaloes with the largest progenitors in the L2
 simulations. Subhaloes with top 100 progenitors are denoted in
 blue; the subset with the 11 largest progenitors are plotted in
 purple.}
\label{MapOrbits2}
\end{figure*}

Aq-A2 exhibits the cleanest structure of all the haloes, with strong
clustering around the pole and antipole, joined by two strands. Aq-B2
is, in contrast, characterised by irregular structures concentrated around regions distant from the main halo spin poles. All of the other haloes exhibit clustering around the main halo
spin, with other, local, features apparent. The densest point vector
position is always closer to the main halo spin than to its antipole.
One may think of Figs.~\ref{MOT1} to \ref{MassBins} as an integration
around lines of equal angle from the red and blue circles. As noted
above, we are particularly interested in those subhaloes that are most
likely to host satellites, and so we repeat this plot for the 100
subhaloes with largest progenitors in Fig.~\ref{MapOrbits2}.

As expected from Fig.~\ref{MassBins}, the 100 subhaloes with the
largest progenitors trace the underlying structure of subhaloes in the
map traced in Fig.~\ref{MapOrbits}. A few of them lie in regions where
there are few subhaloes of any mass, and so we might expect to find
satellite galaxies spatially removed from the disc-of-satellites for at least some portions of their orbits. The
majority, however, lie within underlying structures. The subhaloes with
the top 11 progenitors cluster in the same way as the rest of the
top 100. Thus, we conclude that observed satellite galaxies should
also exhibit coherent motion.

\subsection{The origin of coherent rotation}

\begin{figure*}
  \includegraphics[angle=-90, width=1.0\textwidth]{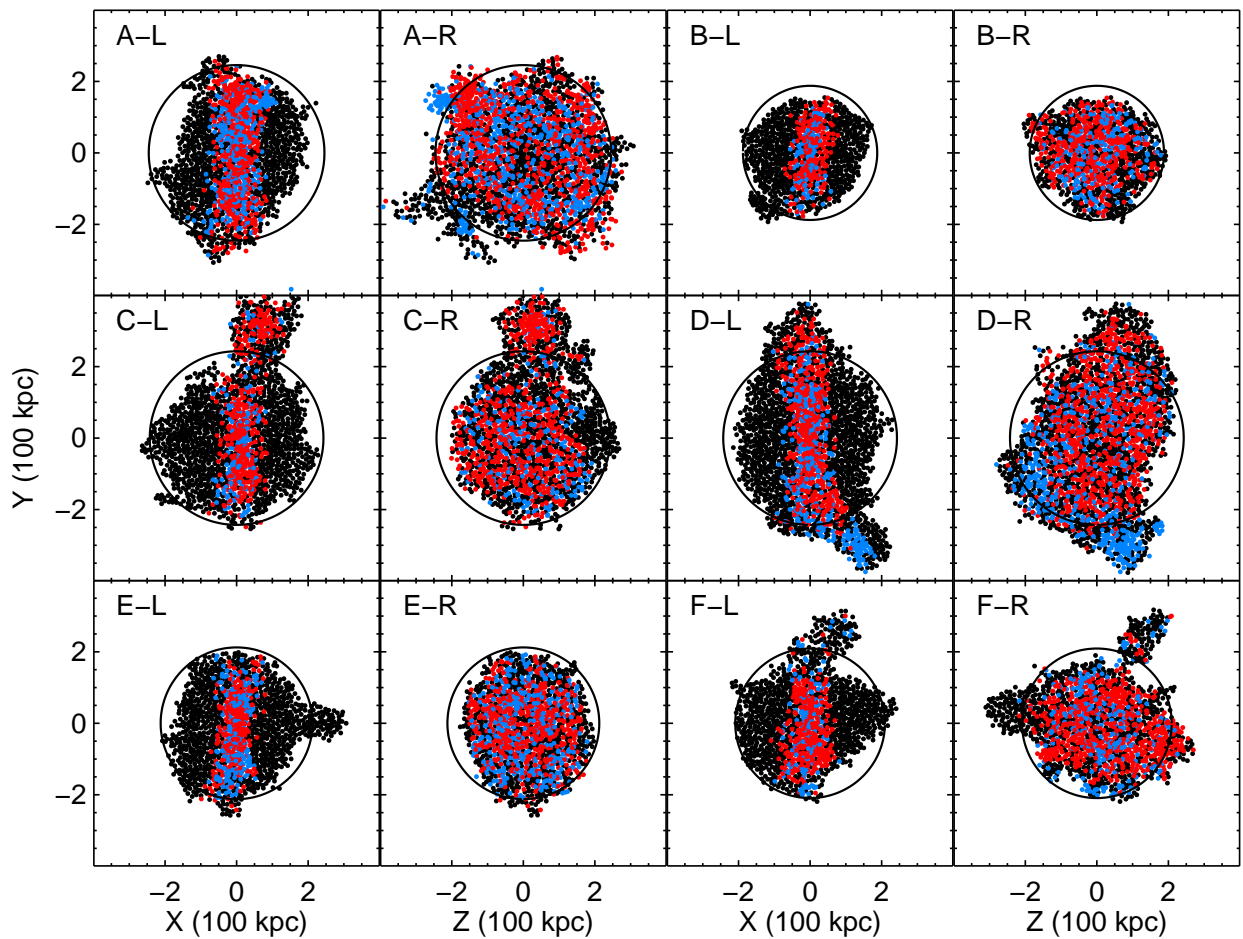}
  \caption{$X-Y$ positions for subhaloes at
  $z=0$ in two projections. Subhaloes with cos~$\theta_{\mathrm{H\cdot S}}$$>0.9$
  are indicated in red and those with cos~$\theta_{\mathrm{H\cdot S}}$$<-0.9$ in
  blue. All other subhaloes are shown in black. The black circle marks
  the virial radius, $r_{200}$. \emph{Left panels (i.e. L):} the $X$
  axis points in the direction of the main halo spin, so those
  subhaloes with orbit vectors parallel and antiparallel to the main
  halo spin appear as a band parallel to the $Y$ axis. \emph{Right Panel (R):}
  looking down the $X$ axis, so the main halo is spinning
  anticlockwise. The red and blue points are plotted in a random order on top of the black.}
\label{DP}
\end{figure*}

The importance of filamentary accretion can be appreciated by
examining the positions of the subhaloes at different snapshots in the
simulation. In Fig.~\ref{DP} we plot the positions of all the
subhaloes present at $z=0$, relative to the centre of the main halo 
in two projections. On the left, the main halo angular momentum vector
points along the positive X-axis, so that the subhalo populations that have 
cos~$\theta_{\mathrm{H\cdot S}}$$>0.9$ (red) and cos~$\theta_{\mathrm{H\cdot S}}$$<-0.9$
(blue) appear as an edge-on thick disc. On the right, the angular
momentum vector points out of the plane of the page.

\begin{figure*}
  \includegraphics[angle=-90, width=1.0\textwidth]{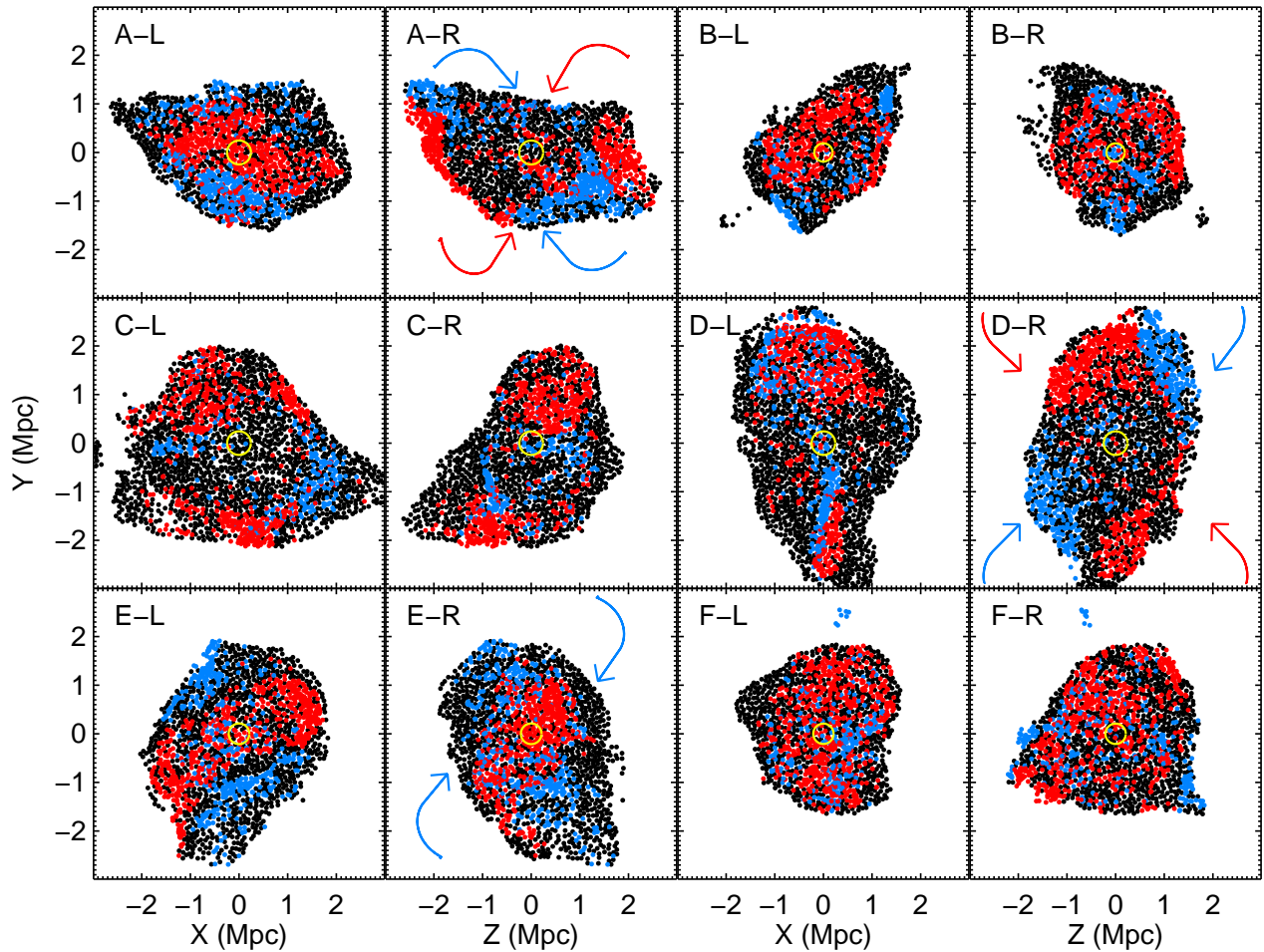}

  \caption{$X-Y$ comoving positions at $z=127$ of the centre of mass of the
    particles that end up in each subhalo at $z=0$. The coordinates are as
    in Fig.~\ref{DP}, with the main halo spin at $z=0$ still determining
    the projections. The final virial radius is indicated in yellow. As stated in the text, the haloes Aq-A2 and Aq-D2 are found to accrete their `red' and `blue' subhaloes in the plane of the main halo spin, and so we have added coloured arrows to the A-R and D-R panels to illustrate schematically the accretion paths for the different subhalo populations. The `blue' subhaloes of Aq-E2 exhibit similar behaviour, and so we have also included arrows to indicate their motion in E-R.}
  \label{DI}
\end{figure*}

In Fig.~\ref{DI} we investigate the origins of the different populations of subhaloes by plotting their positions in the initial conditions. No subhaloes have condensed at this early time, so we define the `position' of each subhalo as the centre-of-mass of all the particles that will be members of that subhalo at redshift zero. Plotting the position of the most-bound particle rather than the centre-of-mass makes no difference to the appearance of the plot, and the plotting procedure followed is exactly the same as that used for Fig.~\ref{DP}.

\begin{figure*}
$\begin{array}{c@{\hspace{-0cm}}c@{\hspace{-0cm}}}
        \includegraphics[angle=-90,width=0.50\textwidth]{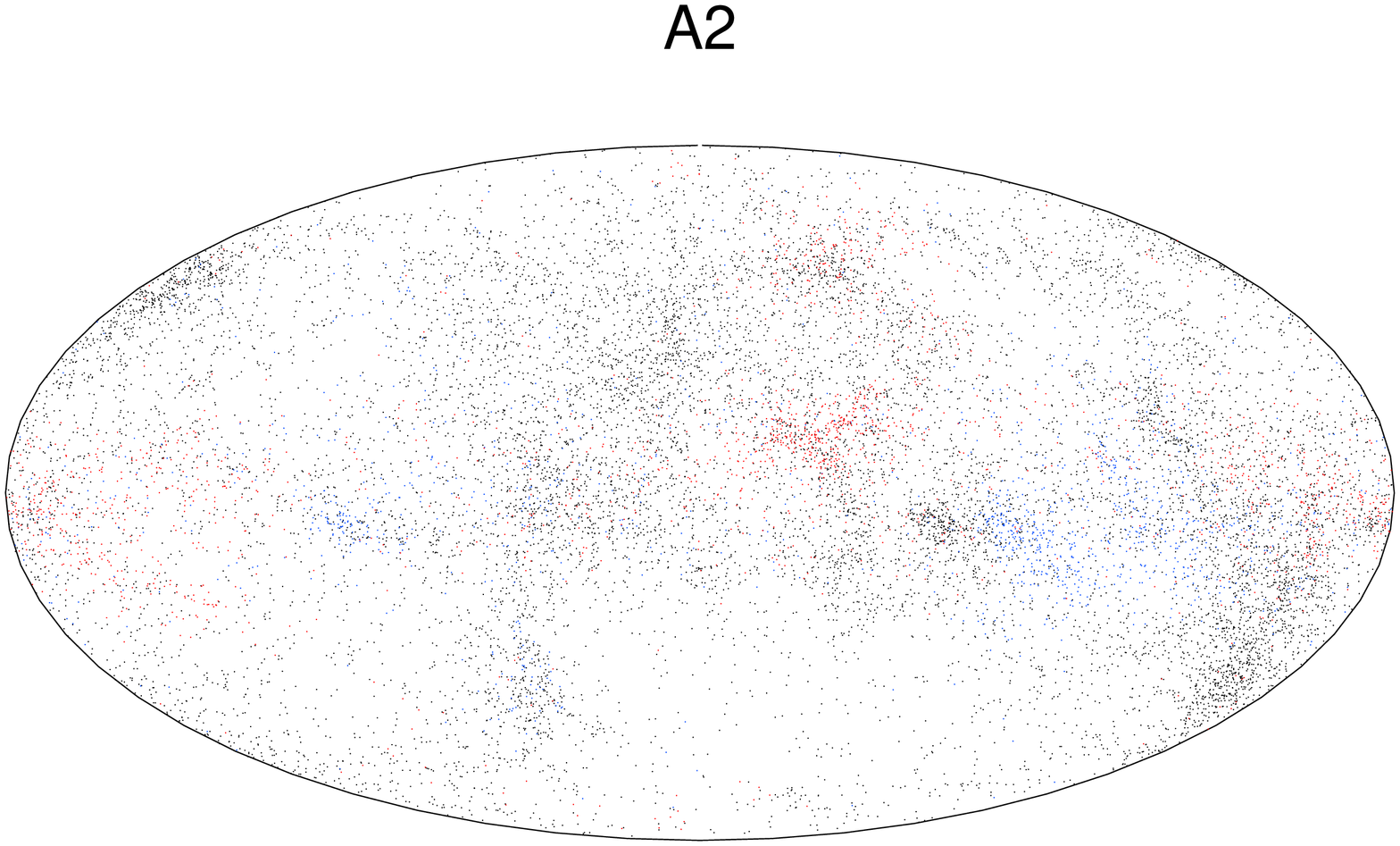} &
        \includegraphics[angle=-90,width=0.50\textwidth]{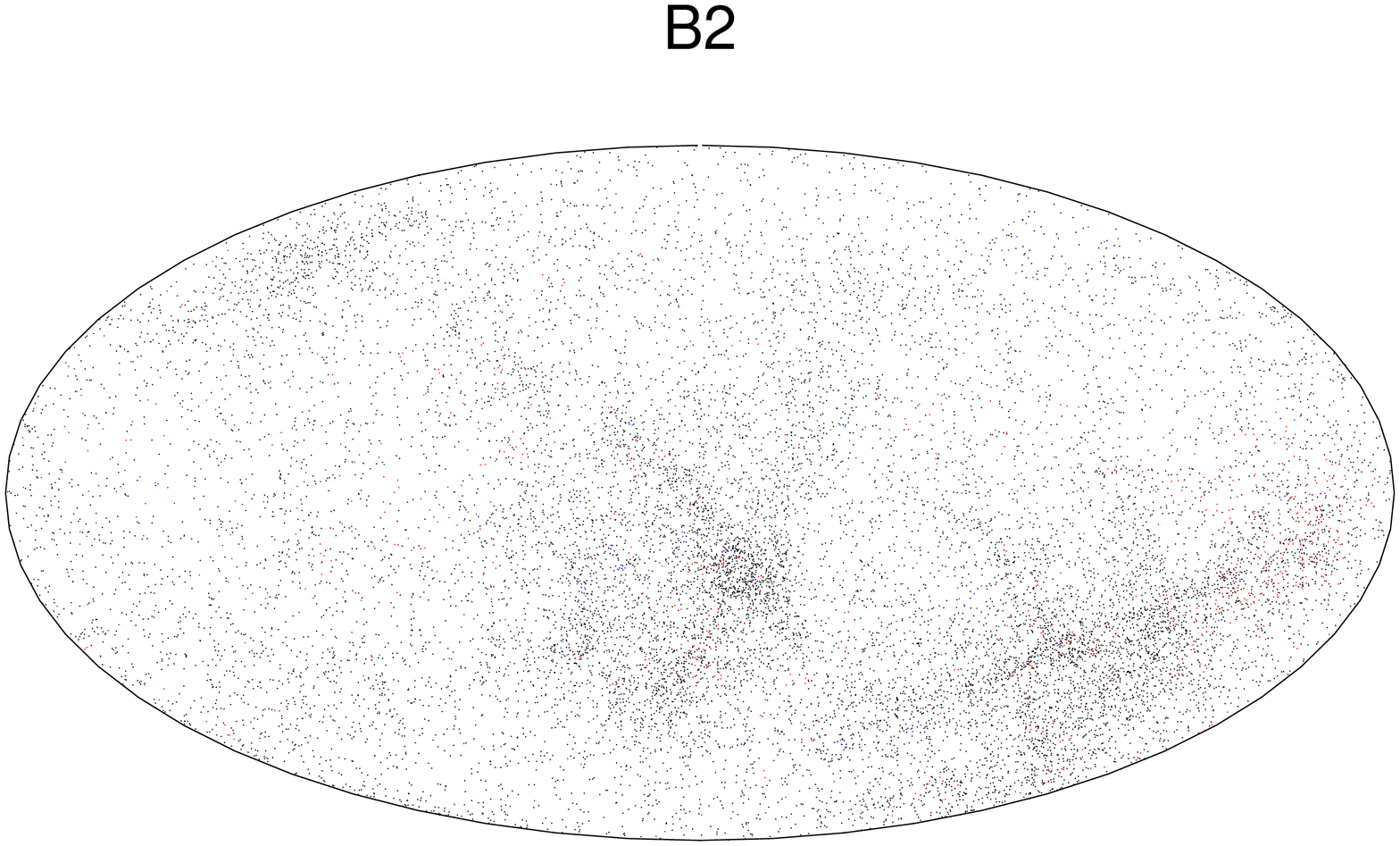}\\
	\includegraphics[angle=-90,width=0.50\textwidth]{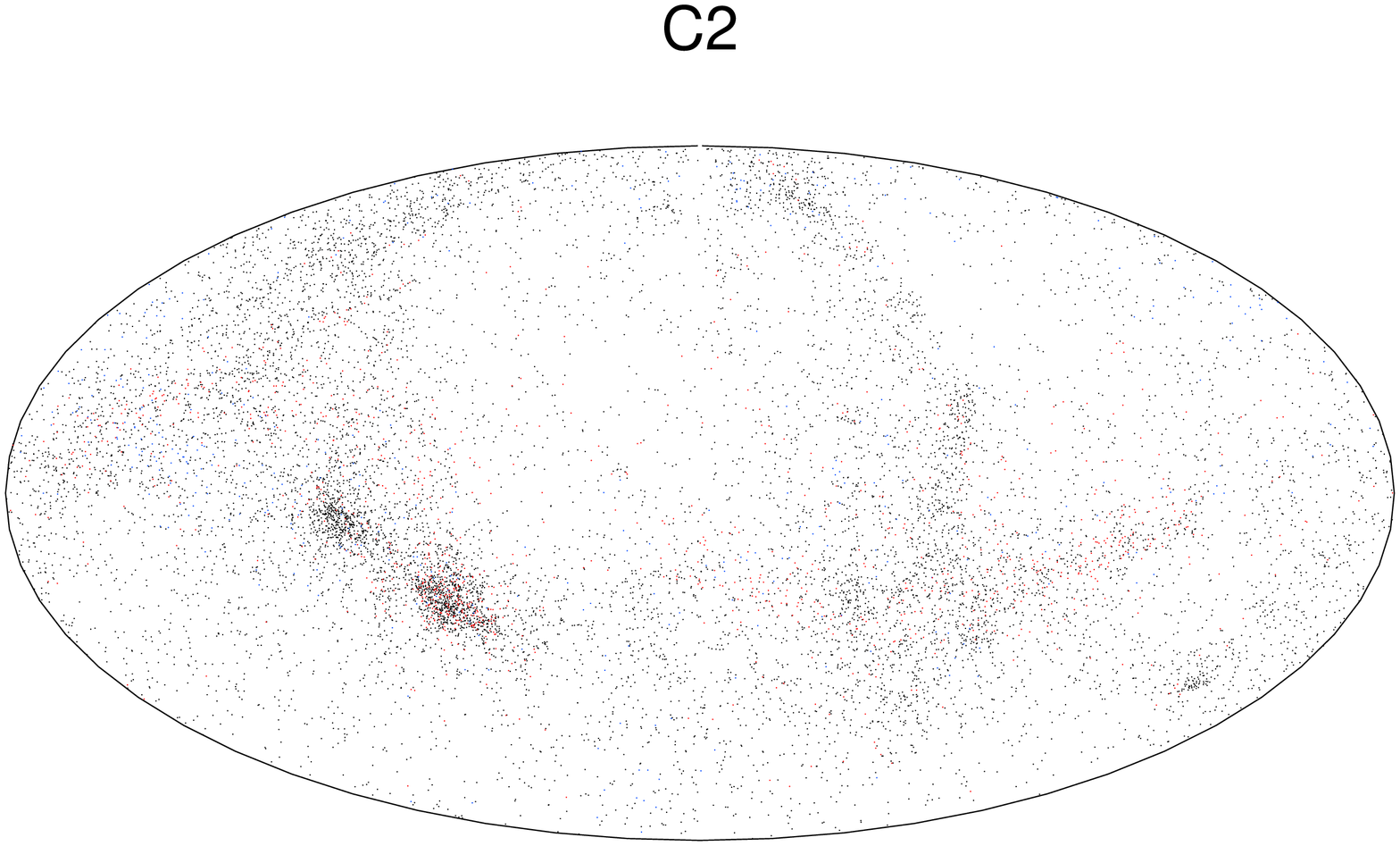} &
        \includegraphics[angle=-90,width=0.50\textwidth]{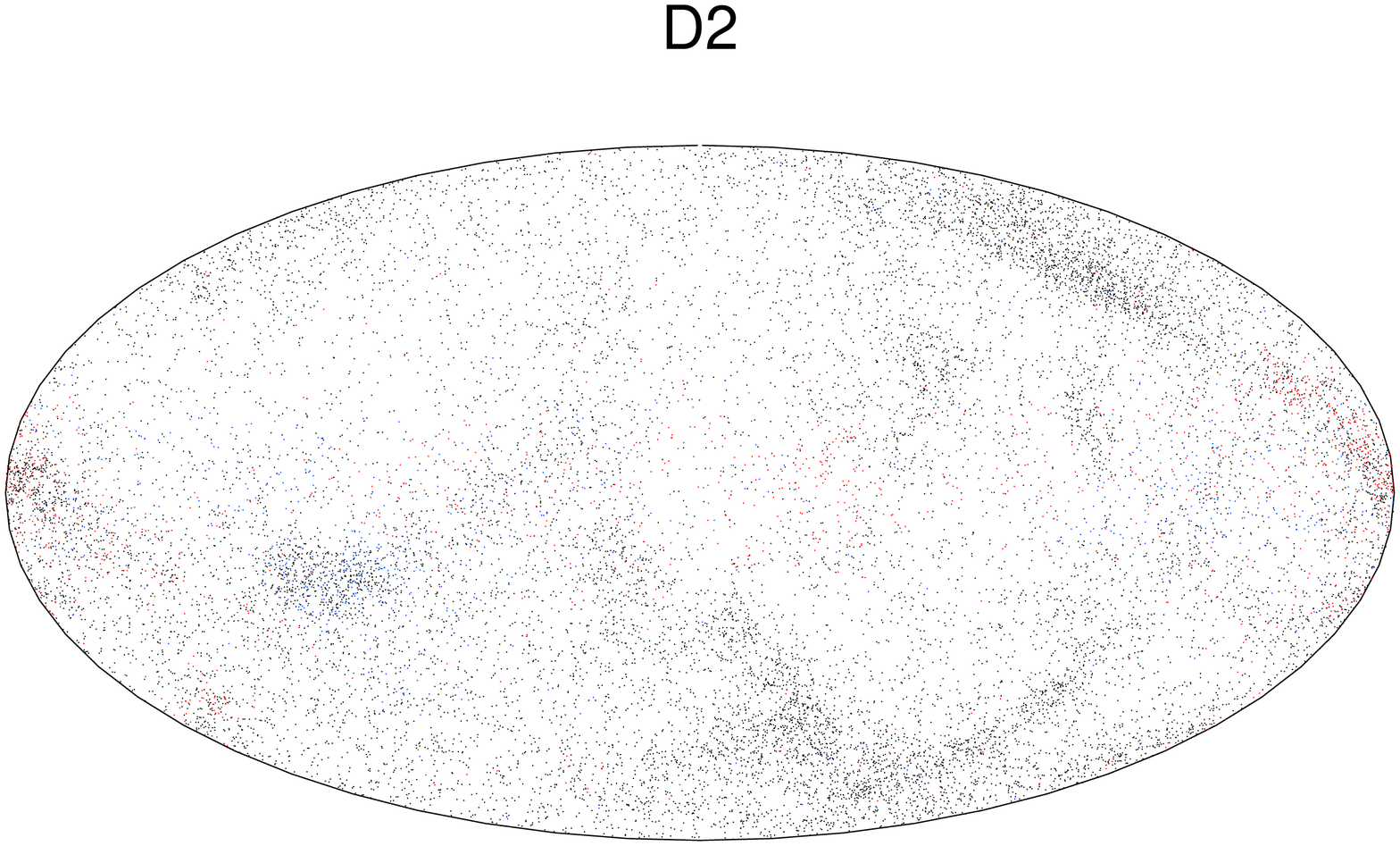}\\
	\includegraphics[angle=-90,width=0.50\textwidth]{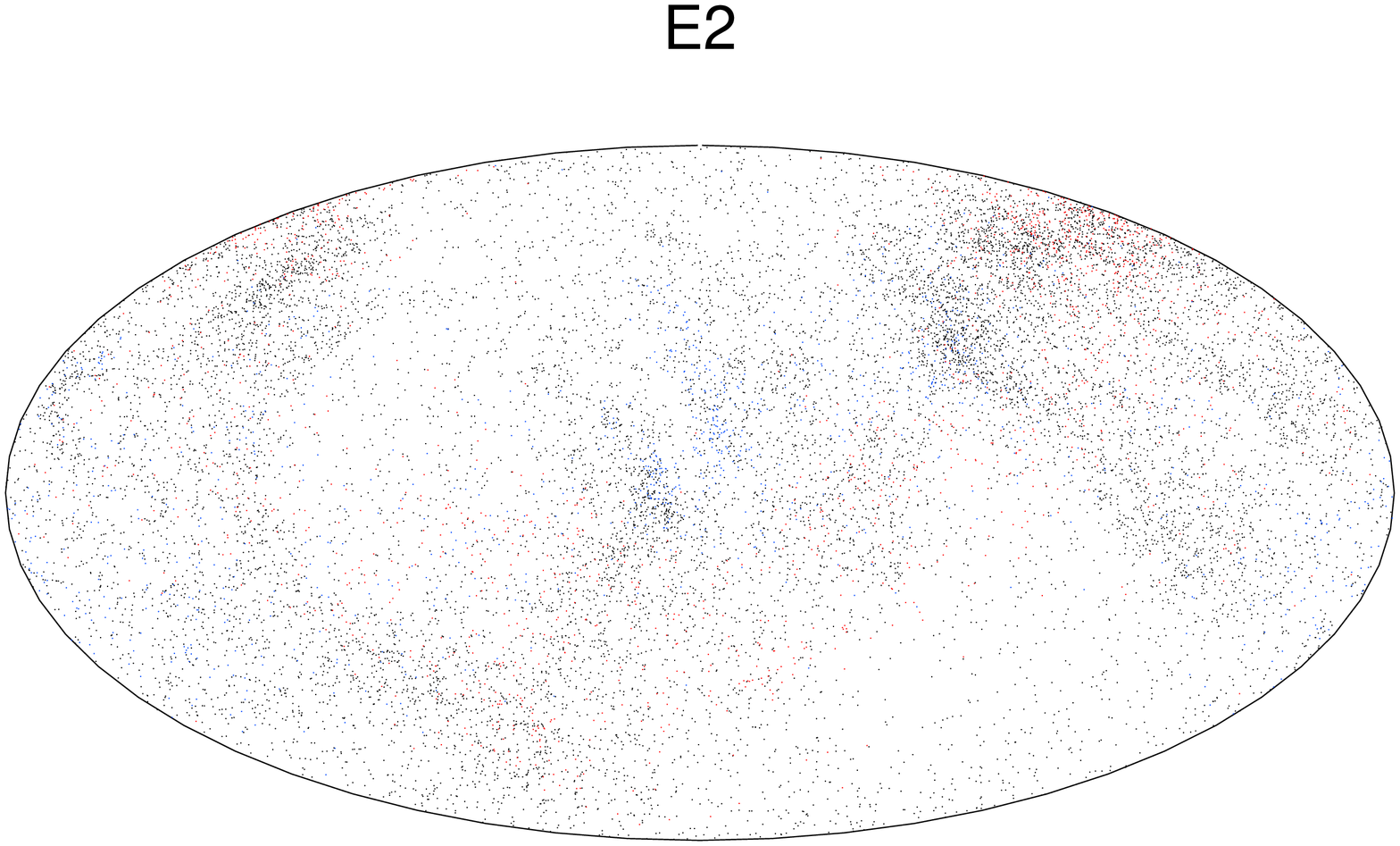} &
        \includegraphics[angle=-90,width=0.50\textwidth]{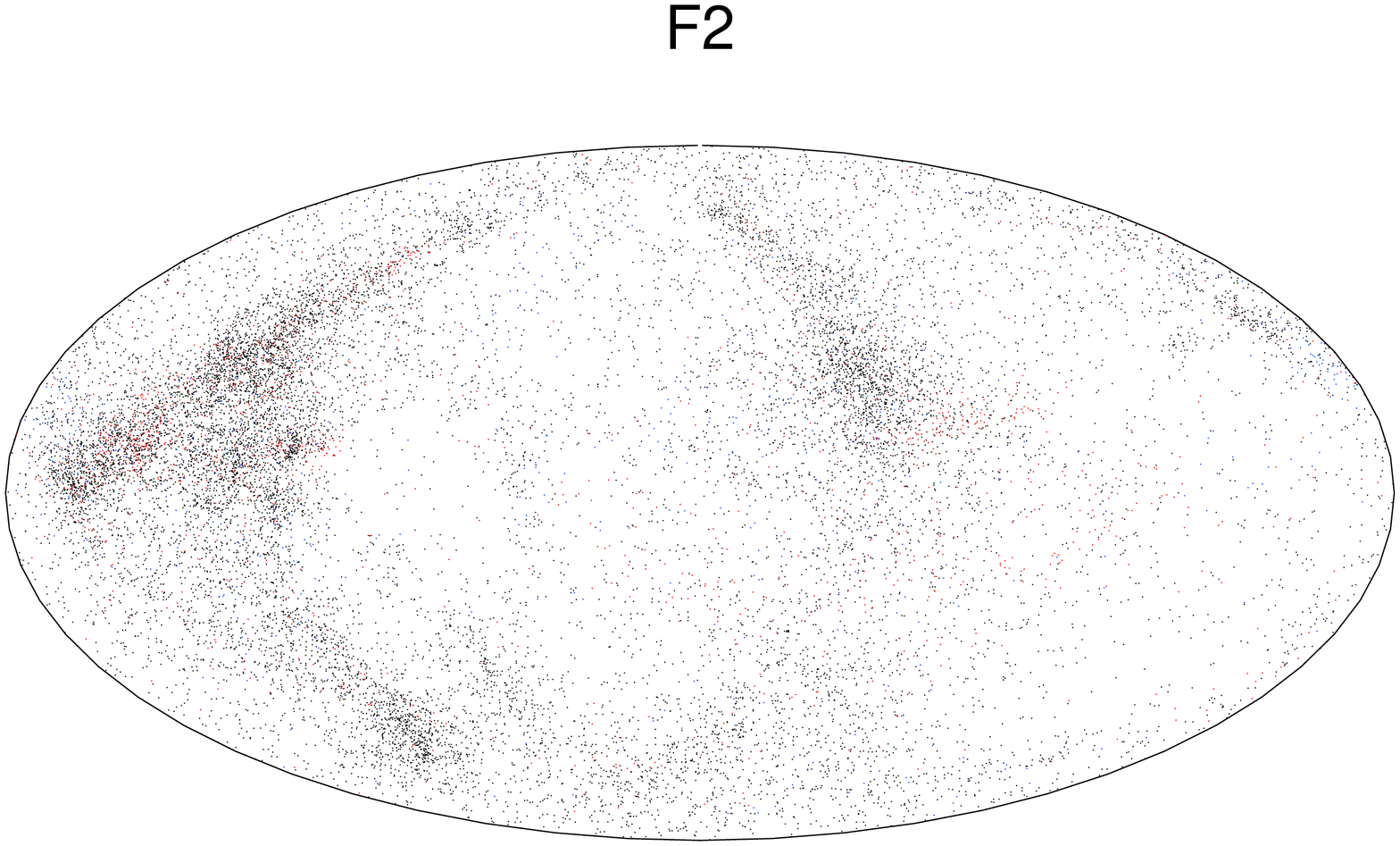}\\
\end{array}$
 \caption{Mollweide projections of the position at which each subhalo
 enters the main halo. Subhaloes that end up in parallel spin orbits
 are shown in red, those that end up in antiparallel spin orbits in
 blue, and those with intermediate orbits in black. The main halo spin
 points towards the north pole of each projection. Higher quality versions of these maps are available at: \emph{http://astro.dur.ac.uk/$\sim$d50wse/page2.html}}
\label{MapPos}
\end{figure*}

All of the haloes that have an excess of near-antiparallel subhaloes in Fig.~\ref{MOT1} show a delineation in the positions of the different subhalo populations. There is also a degree of segregation in Aq-C2, however we find no clear delineation for haloes Aq-B2 and Aq-F2. We can observe how these segregated initial positions evolve into the orbital configurations at the final time by examining snapshots of intermediate redshift. In the cases of haloes Aq-A and Aq-D, we find that, independent of resolution, the motion of interest of these delineated subhaloes occurs within the plane of the main halo spin (the panels A-R and D-R). This enables us to describe this motion simply with the schematic arrows shown in these panels in Fig.~\ref{DI}. In both of these haloes the near-parallel and near-antiparallel populations collapse to form filaments in their segregated regions. Crucially these filaments are not straight, and the subhaloes follow curved paths into the main halo as shown by their same-colour arrows. `Red' (near-parallel) subhaloes will therefore enter the virial radius with an anti-clockwise orbit around the plot centre, whereas the `blue' will adopt a clockwise orbit. In Aq-D the `red' and `blue' filaments are separate entities, but in Aq-A they lie very close together and give the appearance of one filament fed at each end by two `strands'. One of these strands then supplies the near-parallel subhaloes and the other the near-antiparallel.

Aq-E near-antiparallel subhaloes are also accreted through a pair of curved filaments approximately in the plane of the final main halo spin, and so we illustrate the motion of these subhaloes with arrows in Fig.~\ref{DI} panel E-R. By contrast, the accretion of the red subhaloes is more complex and involves motion at a significant angle to the plane of the main halo spin, and for this reason we do not draw the corresponding red arrows. In Aq-C some of the `red' subhaloes do accrete in a filament, but a large proportion end up in the large lump visible at the top of Fig.~\ref{DP} panels C-L and C-R. No coherent inflow pattern is apparent for the small population of `blue' subhaloes.

We can describe the accretion geometry further by determining
where each subhalo enters the main halo. We find the redshift at which
each subhalo attains its highest mass (taken to indicate the time when
it falls into the virial radius of the main halo) and thus determine
its infall position relative to the main halo centre. The results are
plotted in Fig.~\ref{MapPos}, which is oriented such that the main
halo spin points towards the north pole of each projection. We can see
that, independent of resolution, the subhalo populations that end up in parallel and antiparallel
spin orbits in Aq-A, Aq-D, and Aq-E originate from preferential 
directions as expected from our visual examination. A majority of
subhaloes in Aq-D and Aq-A accrete close to the equator, also as
expected, whilst Aq-E acquires a significant proportion of its
parallel orbit subhaloes from a patch of sky close to the main halo
pole. Any demarcation for haloes Aq-B, Aq-C, and Aq-F is less clear,
suggesting that filaments played a lesser role in their accretion
history.

\section{Conclusions}
\label{Co}   

In this paper we have characterized the distribution of subhalo orbits
in the Aquarius simulations of CDM galactic haloes and attempted to
explain the mechanisms that give rise to them.  We find that the
complex accretion patterns that build up a halo result in different
configurations of subhalo orbits, none of which is close to isotropic. Some are
structured in a symmetric way (Aq-A) relative to the spin poles, while others show no strong pattern (Aq-B). In all six haloes we find a large subhalo population
that has coherent rotation aligned with the spin of the main halo, in
agreement with the results of \citet{li09}. In three cases there is,
in addition, a subhalo population that counter-rotates relative to the
main halo. We trace this rather unexpected arrangement back to the
filamentary nature of subhalo accretion. If galaxies tend to rotate in
the same direction as their parent halo \citep{ba05, basm05, be10}, our results show that it is
possible to generate populations of retrograde satellites. Such a population of retrograde
satellites appears to be present in NGC 5084 \citep{ca97} whereas a
population of prograde satellites appears to be present in the Milky
Way \citep{me08}; \citet{hw10} find equal proportions of prograde and retrograde satellites across a sample of 215 systems. 

Halo Aq-A has a particularly concise formation history. This halo
forms from a filament that collapses at early times and is fed by two
strands at either end. A large fraction of the subhaloes that survive
to the present pass though these strands, and are propelled into
either a prograde or retrograde orbit depending on the strand in which
they originated. Aq-D has a similar formation history and outcome,
whereas Aq-E shows that it is possible to end up with a similar orbital
arrangement by a different, more complex path. Coherent rotation is
exhibited by the entire population of subhaloes, not just those 
with the most massive progenitors which according, for example, to
\cite{li05} are the most likely to host visible satellites. 

Our analysis has implications for the expected bulk kinematics of
satellite galaxies which may be probed in future galaxy surveys. We
expect a variety of orbital configurations reflecting the variety of
halo formation histories. Quasi-planar distributions of coherently
rotating satellites should be commonplace, most rotating in the same
direction as the halo (and, by implication, the main galaxy) but some
in the opposite direction as found by \citet{hw10}.

\section*{Acknowledgements}
MRL would like to thank Andrew Cooper and Owen Parry for useful
discussions, and acknowledges a PhD studentship from the Science and
Technologies Facilities Council (STFC). CSF acknowledges a Royal
Society Wolfson Research Merit award. We thank an anonymous referee for helpful comments. This work was supported also by
an STFC rolling grant to the ICC. 


\bsp

\label{lastpage}

\end{document}